\documentclass{emulateapj}
\usepackage{epsfig}
\usepackage{natbib}

\newcommand{\vcn}{v_{\mathrm{circ}}}
\newcommand{\vorb}{v_{\mathrm{orb}}}

\newcommand{\vmax}{v_{\mathrm{max}}}

\newcommand{\mbh}{M_{\mathrm{BH}}}
\newcommand{\msat}{M_{\mathrm{sat}}}
\newcommand{\mhost}{M_{\mathrm{host}}}
\newcommand{\rbh}{r_{\mathrm{BH}}}
\newcommand{\re}{R_e}
\newcommand{\sige}{\sigma_e}

\newcommand{\gsat}{\gamma_{\star, \mathrm{s}}}
\newcommand{\ghost}{\gamma_{\star, \mathrm{h}}}

\shorttitle{Satellite Accretion Onto Massive Galaxies}
\shortauthors{Boylan-Kolchin \& Ma}
\slugcomment{\mnras, in press}
\begin{document}

\title{Satellite accretion onto massive galaxies with central black holes}

\author{Michael Boylan-Kolchin\altaffilmark{1} and Chung-Pei
  Ma\altaffilmark{2}}
\altaffiltext{1}{Department of Physics, University of California,
  Berkeley, CA 94720; mrbk@berkeley.edu} 
\altaffiltext{2}{Department of Astronomy, University of California,
  Berkeley, CA 94720; cpma@berkeley.edu} 

\begin{abstract}
  Minor mergers of galaxies are expected to be common in a hierarchical
  cosmology such as $\Lambda$CDM. Though less disruptive than major
  mergers, minor mergers are more frequent and thus have the potential to
  affect galactic structure significantly. In this paper we dissect the
  case-by-case outcome from a set of numerical simulations of a single
  satellite elliptical galaxy accreting onto a massive elliptical galaxy.
  We take care to explore cosmologically relevant orbital parameters and to
  set up realistic initial galaxy models that include all three relevant
  dynamical components: dark matter halos, stellar bulges, and central
  massive black holes.  The effects of several different parameters are
  considered, including orbital energy and angular momentum, satellite
  density and inner density profile, satellite-to-host mass ratio, and
  presence of a black hole at the center of the host.  Black holes play a
  crucial role in protecting the shallow stellar cores of the hosts, as
  satellites merging onto a host with a central black hole are more
  strongly disrupted than those merging onto hosts without black holes.
  Orbital parameters play an important role in determining the degree of
  disruption: satellites on less bound or more eccentric orbits are more
  easily destroyed than those on more bound or more circular orbits as a
  result of an increased number of pericentric passages and greater
  cumulative effects of gravitational shocking and tidal stripping.  In
  addition, satellites with densities typical of faint elliptical galaxies
  are disrupted relatively easily, while denser satellites can survive much
  better in the tidal field of the host.  Over the range of parameters
  explored, we find that the accretion of a single satellite elliptical
  galaxy can result in a broad variety of changes, in both signs, in the
  surface brightness profile and color of the central part of an elliptical
  galaxy.  Our results show that detailed properties of the stellar
  components of merging satellites can strongly affect the properties of
  the remnants.
\end{abstract}

\keywords{galaxies: elliptical and lenticular, cD --- galaxies:
  evolution --- galaxies: formation}

\section{Introduction}
There now exists a strong body of evidence that luminous elliptical
galaxies are, in many ways, a distinct group of objects when compared to
lower-luminosity elliptical galaxies.  This dichotomy is seen not only in
their global structure -- for example, isophotal shapes
\citep{kormendy1996} and degree of rotational support \citep{davies1983,
  bender1992} -- but also in their central properties: giant ellipticals
($M_v<-22$) have shallow central surface brightness profiles ($-d\ln
\Sigma/d\ln R < 0.26$), little rotation but velocity anisotropy, and boxy
isophotal shapes, whereas spiral bulges and fainter ellipticals
($M_v>-20.5$) exhibit central power-law surface brightness profiles ($-d\ln
\Sigma/d\ln R > 0.5$), significant rotational support, and disky isophotal
shapes \citep{lauer1995, faber1997, ravindranath2001, lauer2005,
  ferrarese2006}.

Different environments and formation histories are plausibly responsible
for these two classes of galaxies.  The fainter ellipticals are found in
all environments including the field; their dense power-law centers may be
a signature of dissipation in gas-rich mergers of disk galaxies (e.g.,
\citealt{faber1997, genzel2001}).  The luminous ellipticals, on the other
hand, are predominately found in galaxy clusters, some as brightest cluster
galaxies and cD galaxies at or near the centers of clusters.  Their shallow
central luminosity profiles and low-density cores appear particularly
difficult to form and preserve in theoretical models.

Black holes provide an attractive model for the initial formation of the
cores: mergers of two cuspy ($\rho \propto r^{-2}$) stellar bulges with
central black holes result in the sinking of the black holes to the center
of the merger remnant by dynamical friction and the formation of a black
hole binary \citep{begelman1980}.  This process, and the subsequent orbital
decay (``hardening'') of the binary, result in energy transfer from the
black holes to the stellar system, and the combined effect is to reduce the
initial cusp to a profile that is nearly cored in projection
\citep{ebisuzaki1991, quinlan1997,milosavljevic2001}.

How the shallow surface brightness profiles of luminous ellipticals evolve
and are preserved (for the most part) in subsequent mergers, however, is an
open question and is the focus of this paper.  The stellar mass within the
core radius (defined to be the radius where the local logarithmic slope of
the surface brightness is $-0.5$) is of order 1\% or less of the total
stellar mass; a small amount of mass deposit or removal can therefore
easily alter the inner profiles.  To maintain such shallow stellar
profiles, massive elliptical galaxies, once formed, must somehow prevent
additional stellar mass from being brought into their central regions
through the frequent satellite accretion that is a hallmark of cold dark
matter-based cosmological models.  In particular, cosmological $N$-body
simulations that model only dark matter have revealed a rich spectrum of
dense dark matter subhalos, some of which can survive complete tidal
disruption and orbit within their host halos.  A logical expectation is
that some of the massive core ellipticals have accreted at least one
lower-mass galaxy with cuspier central density over their lifetimes.  It is
therefore intriguing that surface brightness cores in luminous ellipticals
are ubiquitous.

In this paper, we report the results from a set of merger simulations that
explore the effects of a single satellite accreting onto a massive
elliptical galaxy.  Our simulations use realistic galaxy models composed of
all three components that are dynamically relevant for elliptical galaxies:
dark matter halos, stellar spheroids, and central supermassive black holes
(section~\ref{sec:sims}).  As we show, each component plays an important
role in the outcome of the accretion events and therefore should be
included.  The dark matter halos set the gravitational potential well on
large scales and determine the orbital evolution in the early stages of the
merger (before the stellar bulges interact strongly).  The bulges dominate
the gravitational potential on smaller ($\sim$ kpc) scales and thus affect
the orbital evolution at late times in the merger.  On sub-kpc scales,
supermassive black holes can dominate the potential well and thus
contribute to the structural evolution of satellites, sometimes quite
strongly.  Even though the simulations are dissipationless, the three
components exchange energy gravitationally with one another during the
merger, leading to observable effects on the stellar component that would
not be fully captured if any component were neglected.

An additional feature of our simulations is that we choose the orbital
parameters based on distributions of orbits determined from large
cosmological dark-matter-only simulations.  By applying a semi-analytic
prescription of dynamical friction and tidal stripping (section~2.3), we
are able to start our simulations not at the virial radius of the host,
where the cosmological distributions of halo merger orbits are determined,
but at the smaller scale of four effective radii of the host's bulge.

Our analysis focuses broadly on two aspects of satellite accretion: orbital
decay and structural evolution of the satellite during the merger process
(section~\ref{sec:fate}) and observable signatures of infalling satellites
on the host galaxies (section~\ref{sec:observables}).  For the former, we
investigate the dependence of an accreting satellite's fate on the initial
orbital energy and angular momentum (section~3.1), the satellite's density
(section~3.2), the presence of central black holes (section~3.3) and dark
matter halo (section~3.4), and other factors such as the satellite-to-host
mass ratio and the inner density profile of the satellite (section~3.5).
For the observable consequences of satellite accretion, we investigate a
number of different properties including surface brightness profiles
(section~4.1), colors and color gradients (section~4.2), and mass deposit
at large radii and kinematic structures (section~4.3).

\section{Numerical simulations}
\label{sec:sims}
\begin{deluxetable}{cccccccc}
\tablecaption{Summary of Satellite Accretion Simulations
\label{table-ICs}}
\tablehead{
\colhead{\textbf{Run}}
& \colhead{$\mathbf{N_p}$} 
& \colhead{$\mathbf{v_t / v_r}$}
& \colhead{$\mathbf{\frac{v_{\mathrm{orb}}}{\vcn}}$}
& \colhead{$\gsat$}
& \colhead{\textbf{BH?}} 
& \colhead{$\mathbf{\re}$/kpc} 
& \colhead{\boldmath{$\epsilon$/pc}}\\
(1) & (2) & (3) & (4) & (5)  & (6) & (7) & (8)
}
\startdata
\cutinhead{Three-component (dark matter+star+black hole) Runs}
H1 & $1.1\times 10^6$ & 0 & 1.5 & 1.0 & host & 1.39 & 33\\
H1hi & $5.5 \times 10^6$ & 0 & 1.5 & 1.0 & host & 1.39 & 20\\
H1n& $1.1\times 10^6$ & 0 & 1.5 & 1.0 & no & 1.39 & 33\\
H1s & $1.1\times 10^6$ & 0 & 1.5 & 1.5 & host & 1.39 & 33\\
L1 & $1.1\times 10^6$ & 0 & 0.75 & 1.0 & host & 1.39 & 33\\
H2 & $1.1\times 10^6$ & 0.5 & 1.5 & 1.0 & host & 1.39 & 33\\
H2n & $1.1\times 10^6$ & 0.5 & 1.5 & 1.0 & no & 1.39 & 33\\
H2d & $1.1\times 10^6$ & 0.5 & 1.5 & 1.0 & host & 1.07 & 33\\
H2dd & $1.1\times 10^6$ & 0.5 & 1.5 & 1.0 & host & 0.823 & 33\\
H2ddd & $1.1\times 10^6$ & 0.5 & 1.5 & 1.0 & host & 0.700 & 33\\
H3 & $1.1\times 10^6$ & 1.0 & 1.5 & 1.0 & host & 1.39 & 33\\
H3s & $1.1\times 10^6$ & 1.0 & 1.5 & 1.5 & host & 1.39 & 33\\
M3 & $1.1\times 10^6$ & 1.0 & 1.1 & 1.0 & host & 1.39 & 33\\
L3 & $1.1\times 10^6$ & 1.0 & 0.75 & 1.0 & host & 1.39 & 33\\
L3d & $1.1\times 10^6$ & 1.0 & 0.75 & 1.0 & host & 1.07 & 33\\
H4 & $1.1\times 10^6$ & 2.0 & 1.5 & 1.0 & host & 1.39 & 33\\
H4s & $1.1\times 10^6$ & 2.0 & 1.5 & 1.5 & host & 1.39 & 33\\
L4 & $1.1\times 10^6$ & 2.0 & 0.75 & 1.0 & host & 1.39 & 33\\
&&&&&&\\
\cutinhead{Two-component (star+black hole) Comparison Runs}
T1 & $5 \times 10^5$ & 0 & 1.0 & 1.0 & host & 1.39 &  20\\
T2\tablenotemark{a} & $5 \times 10^5$ & 0 & 1.0 & 1.0 & host & 0.708 & 20\\
T3s\tablenotemark{a} & $5 \times 10^5$ & 0 & 1.0 & 1.5 & host & 1.00 & 20\\
T4 & $5 \times 10^5$ & 0 & 1.0 & 1.0 & host & 0.700 & 20\\
T5 & $5 \times 10^5$ & 0 & 1.0 & 1.0 & both & 0.700 & 20\\
T6 & $5 \times 10^5$ & 0 & 1.0 & 1.0 & both & 0.700 & 3.3\\
T7 & $5 \times 10^5$ & 0.25 & 1.0 & 1.0 & host & 1.39 & 20\\
T8 & $5 \times 10^5$ & 0.5 & 1.0 & 1.0 & host & 1.39 & 20\\
\enddata
\tablenotetext{a}{satellite-to-host mass ratio of 0.03 rather than 0.1}
\tablecomments{Description of columns:\\
(1) Name of run in order of increasing orbital angular momentum. H,M,L
    indicate high, medium, low orbital energy; n indicates no black holes;
    d indicates denser satellites; s indicates steeper inner density slope
    for stellar component of satellite\\
(2) Total number of simulation particles in the host galaxy (before
    truncation).  The total dark matter-to-stellar mass ratio is 10:1  for
    each galaxy (host and satellite) in the 3-component runs\\
(3) Ratio of initial tangential to radial orbital velocity ($v_t/v_r$) of satellite\\
(4) Ratio of initial orbital velocity to local (host) circular velocity\\
(5) Inner slope of the satellite's stellar density profile\\
(6) Which of progenitors has a black hole\\
(7) Effective radius $\re$ of the satellite's stellar bulge (in kpc)\\
(8) Spline force softening for stellar, dark matter, and black hole
    particles (in parsecs)}
\end{deluxetable}

\subsection{Initial Galaxy Models}
\label{subsec:params}

The initial host galaxies in our main set of simulations (see Table~1)
contain three components: dark matter halo, stellar bulge, and a central
supermassive black hole, all assumed to be in mutual equilibrium in the
total gravitational potential.  We also perform comparison runs (a) without
black holes to quantify the degree of satellite destruction due to the
presence of the hole; (b) with two black holes, one at the center of each
merging galaxy to study the effects due to (unequal mass) black hole
binaries; and (c) without a dark matter halo to quantify the differences in
satellite orbit and fate between the more realistic three-component runs
and earlier work that modeled only stellar bulge and ignored dark matter.
Luminous galaxies have non-negligible amounts of dark matter on scales of
the half-mass radius (e.g., \citealt{Padmanabhan2004}), which can lead to
non-negligible effects on merging satellites.  The initial setup for each
of the three components is discussed below.

\subsubsection{Dark matter halos}

The initial dark matter halos and stellar bulges in our simulations
both have ``$\gamma$ profiles'' \citep{dehnen1993, tremaine1994}: for
component $i$ (either dark matter or stars), the density profile is given by
\begin{equation}
\rho_i(r)=(3-\gamma_i) \frac{M_i}{4 \pi a_i^3} \left(\frac{r}{a_i}
\right)^{-\gamma_i} \left(1+\frac{r}{a_i}\right)^{-(4-\gamma_i)} \,.
\label{eqn:gamma}
\end{equation}
There are three parameters in this model: $\gamma$ controls the inner
density slope, $a$ is a characteristic radius, and $M$ is the total mass of
the given component.  Additionally, we impose an outer density cut-off,
mimicking the effects of tidal truncation.  Details of this truncation and
several other properties of these models are provided in the Appendix.

We assign the dark matter halos $\gamma_{\mathrm{DM}}=1$ and match the
inner density of this \citet{hernquist1990} profile to that of the
\citet[NFW]{navarro1997} profile having the same mass within its virial
radius (see \citealt{springel2005a} for details of this procedure).  Since
the outer regions of our halos are truncated, the mass distribution on
important scales is virtually identical to NFW profiles.  The mass and
radius in our non-cosmological simulations can be rescaled arbitrarily, but
for better physical insight, we choose to quote all the results in this
paper for a fiducial (untruncated) dark matter halo mass and radius of
($M_{DM}, a$)=($3.33 \times 10^{12} M_{\odot},\, 33$ kpc) and ($3.33 \times
10^{11} M_{\odot}, \, 13.33$ kpc) for the host and satellite, respectively.
We generally consider 1:10 mergers in this paper since this mass ratio is
small enough be in the minor merger regime but large enough for mergers to
occur on reasonably short timescale (see Sec~2.3 below and, e.g.,
\citealt{taffoni2003} for calculations of merger timescales for a variety
of mass ratios).  We also investigate 1:33 mergers to check the robustness
of our predictions (see Sec~3.5).

\subsubsection{Stellar bulges}

\begin{figure*}
  \centering
  \includegraphics[scale=0.8]{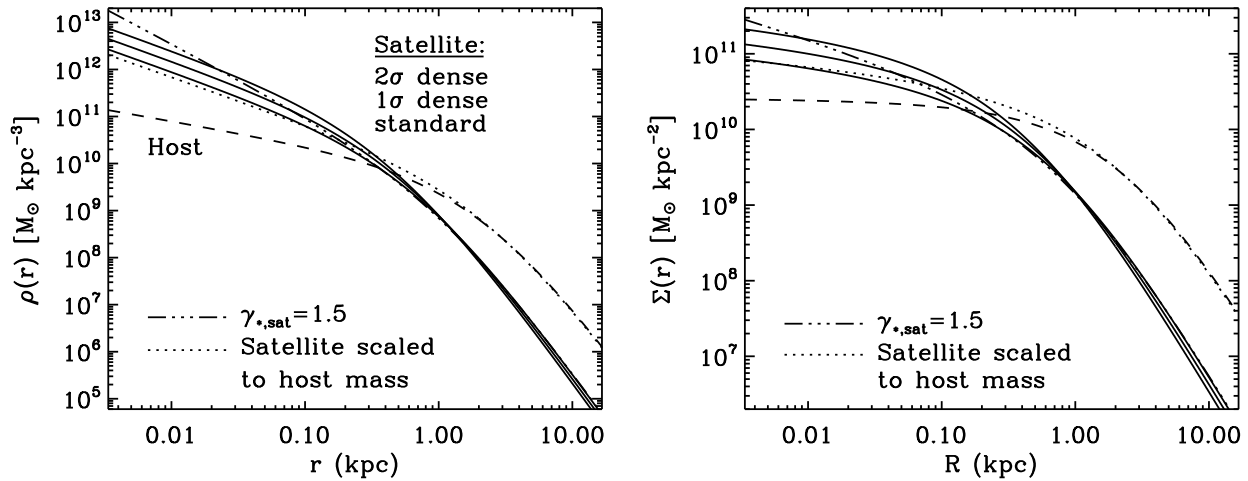}
  \caption{ Initial galaxy models for the simulations.  Left: stellar
    mass density profiles of $\gsat=1$ (set of solid curves) and
    $\gsat=1.5$ (dot-dot-dot-dashed curve) satellites.  Also shown is
    the initial host stellar density profile (dashed curve) as well as
    the host profile predicted by the SDSS $\re-M_\star$ relation if
    the host had a Hernquist density profile (dotted curve).  Right:
    surface mass density profiles corresponding to the curves on the
    left.}
  \label{fig:ICs}
\end{figure*}

The stellar bulges are self-consistently embedded in dark matter halos with
mass $M_{DM}=10 \,M_{\star}$ before truncation.  For the host galaxies, the
initial stellar bulges are all given $\ghost=0.5$ to model the nearly flat
central surface brightness profiles of massive ellipticals (dashed curves
in Fig.~\ref{fig:ICs}).  To assign the scale radius $a$ in
equation~(\ref{eqn:gamma}) for a given stellar mass $M_\star$, we use the
mean size-mass relation, $\re \propto M_{\star}^{0.6}$, for early-type
galaxies \citep{shen2003} determined from the Sloan Digital Sky Survey
\citep[SDSS;][]{york2000} and the stellar mass estimates of
\cite{kauffmann2003}.  Note that the relation between the effective radius
$\re$ and scale radius $a$ depends on $\gamma$: $\re/a=(2.358,\, 1.815,\,
1.276, 0.744)$ for $\gamma=(0.5,\, 1.0,\, 1.5,\, 2.0)$.  For our fiducial
mass scale, the stellar mass of the host is $M_{\star}=3.3 \times 10^{11}
\, M_{\odot}$, which corresponds to a galaxy of magnitude $M_v \approx -22$
(e.g., \citealt{gebhardt2003}).  For $\ghost=0.5$, the corresponding
effective radius is $\re=5.5$ kpc ($a=2.33$ kpc), and the core radius is
approximately 400 pc, near the upper end of the observed distribution of
core sizes at our chosen stellar mass \citep[Fig.~5]{lauer2006}.

For the satellite galaxies, the initial stellar bulges are given
$\gsat=1.0$, resulting in initial surface brightness profiles (solid
curves in Fig.~\ref{fig:ICs}) that closely approximate the $r^{1/4}$ law
\citep{de-vaucouleurs1948}\footnote{The Hernquist ($\gamma=1$) profile
  actually deviates from the $r^{1/4}$ law at both small and large radii
  \citep{hernquist1990, boylan-kolchin2005} but is a good match to the
  $r^{1/4}$ profile for the
  region containing most of the mass.}.  We
also perform comparison simulations with $\gsat=1.5$ satellites (keeping
$\re$ fixed) and find increasing $\gsat$ to be somewhat degenerate with
making the satellites more compact by reducing $a$ (see
Sec.~\ref{sec:mass_gamma}).  Since tidal effects depend on the local ratio
of the satellite and host densities, we find it important to take into
account the scatter in the SDSS $\re-M_{\star}$ relation, which has a
dispersion of $\sigma_{\ln R} \approx 0.26$ for early-type galaxies
\citep{shen2003}.  For satellites of a given $M_{\star}$, we therefore
consider galaxies on the SDSS $\re-M_\star$ relation as well as those with
$\re$ that are $1\sigma$ and $2\sigma$ smaller, referring to them as
``standard'', ``$1\sigma$ dense'', and ``$2\sigma$ dense'' satellites in
subsequent discussions.

Fig.~\ref{fig:ICs} shows the stellar density (left panel) and surface
brightness (right panel) profiles of our initial galaxy models for the
host (dashed curves), the $\gsat=1.0$ satellites (solid curves), and the
$\gsat=1.5$ satellite
(dot-dot-dot-dashed curve).  The
satellites are denser than the host for $r \la 300$ pc, while the surface
brightness of the satellites exceed that of the host galaxy for $R \la
150-300$ pc (and only by a factor of 3 at 10 pc for the standard
satellite).  Fig.~\ref{fig:ICs} also shows the stellar density profile for
a galaxy with the same stellar mass as the host, assuming $\ghost=1$ and
$\re$ following the SDSS relation (dotted line).  Its inner density is
nearly identical to that of the satellite, which underlines the distinction
between core and power-law galaxies at fixed luminosity.

The line-of-sight stellar velocity dispersions are 240 km s$^{-1}$ for the
host and 140 km s$^{-1}$ for the (standard) satellite, placing them on the
Faber-Jackson \citeyearpar{faber1976} and fundamental plane
\citep{dressler1987, djorgovski1987} relations.

\subsubsection{Central black holes}

We model the black hole as a point mass with $\mbh/M_{\star}=2 \times
10^{-3}$, similar to the relation observed in galactic nuclei
\citep{magorrian1998, ferrarese2000, gebhardt2000, haring2004}.  The black
hole exerts a dynamical influence over the stars (and dark matter) within
its sphere of influence, $\rbh$, which we define as the radius enclosing a
mass in stars equal to twice that of the black hole:
$M_{\star} (< \rbh)=2\,\mbh$.
With this definition, the standard $\rbh=G \mbh/\sigma^2$ holds true for the
isothermal sphere.  For reference, $\rbh/a =(0.123,\, 0.0675,\, 0.0258,\,
0.00402)$ for $\gamma=(0.5, \, 1.0,\, 1.5,\, 2.0)$, assuming
$\mbh/M_{\star}=2 \times 10^{-3}$.  Note this scale can be much larger than
the $\approx$ pc scale often attributed to black holes: for our choice for
host galaxy of $a=2.33$ kpc, $\ghost=0.5$, and $M_{\star}=3.3 \times
10^{11} \, M_{\odot}$, we find $\rbh=290$ pc, comparable to the size of the
surface brightness core in our galaxy model (dashed line in the right panel
of Fig.~\ref{fig:ICs}).  In comparison, a profile with $\ghost=2$ but with
the same $M_{\star}$ and $\re$ as our $\ghost=0.5$ profile has $\rbh=7.7$
pc, showing how sensitive the black hole's sphere of influence is on the
stellar density profile under consideration.

For cases with two black holes, another relevant scale is $r_{\rm hard}
=G\mu/4\sigma^2$, which is the separation at which the two black holes form
a hard binary.  This is the radius where the gravitational binding energy
of the binary exceeds the kinetic energy of the stellar background.
Note that for an equal-mass binary in an isothermal stellar background,
$r_{\rm hard}=\rbh/8$, and this is always an upper limit for the cases we
are considering.

\subsection{Methodology and Numerical Effects}
\label{subsec:methods}
All simulations presented are performed using the parallelized tree code
{\sc gadget-2} \citep{springel2005} in pure $N$-body mode.  The initial
positions for the dark matter and star particles are drawn from their
respective spherically symmetric density profile described in Sec.~2.1.
The initial velocities for each component are drawn from that component's
distribution function, assuming the system to be spherical and isotropic
(see \citealt{boylan-kolchin2005} for details).  In using this procedure,
we make no simplifying assumptions about the velocity structure of the
model components and the resulting initial conditions are initially in an
equilibrium that is modified only by numerical effects.  An example of the
numerical stability of our galaxy models is provided in the Appendix.

We have performed resolution tests to determine if the results reported in
Sec.~3 and 4 are sensitive to the number of particles, force resolution, or
force softening used in the simulations.  One such test (run H1hi in
Table~\ref{table-ICs}) uses 5 times more particles for both the bulge and
halo and a force softening that is a factor $5^{1/3}$ smaller than that of
the standard runs.  No differences are found between this run and the
equivalent run (H1) at our standard resolution.  All runs have equal
particle masses for the stellar and dark matter particles to avoid
numerical relaxation; accordingly, the force softening is equal for both
particle species in all runs.  We use the same force softening for the
black hole as well.  In all production runs, energy is conserved to better
than 0.2\%.

For simulations with central black holes, gravitational brownian motion can
be significant in our isolated galaxy models.  The expectation is that the
black hole and star particles will scatter off of each other,
redistributing the particle energies toward equipartition: $(v_{BH}/
\sigma_{\star})^2=3 N_\star^{-1} M_{\star}/ \mbh$ (though see
\citealt{chatterjee2002} for a more complete discussion).  This velocity
scale for reasonable particle numbers is not insignificant: for
$\mbh/M_{\star}=2 \times 10^{-3}$ and $N_{\star}=10^5$, then $v_{BH}/
\sigma_{\star}=0.122$.  Note that the characteristic scale of the
wandering, $r_w \approx v_{BH}/(G \, \overline{\rho})^{1/2}$, can be
somewhat large for a low-density core.  We do observe wandering in our
simulations, but the phase space structure is not affected in any important
way above our force softening scale.

\subsection{Choice of orbital parameters}

In order to make meaningful predictions about satellite accretion, we would
like to use realistic orbital properties drawn from merger events in
cosmological simulations.  This task is non-trivial, however, since mergers
of dark matter halos and mergers of galactic bulges are not the same thing:
analyses of cosmological simulations generally compute merger orbital
properties at a galaxy's dark matter virial radius (e.g.,
\citealt{benson2005, khochfar2006}), whereas we are mainly interested in the
detailed dynamics among dark matter, stars, and black holes after the two
merging galaxies' stellar bulges begin to interact.  It has long been known
that galactic cannibalism may be important for the growth of central
cluster galaxies \citep{ostriker1975,hausman1978,lauer1988} but also must
be relatively inefficient due to the long merging timescales for satellites
\citep{merritt1984,merritt1985}.  As a result, there is a selection effect
favoring satellites with higher masses that have correspondingly shorter
dynamical friction timescales.  For the same reason, accretion events that
lead to merging with a central galaxy should be skewed towards lower
angular momentum orbits than the average orbit for all halos accreted at
$R_{\rm vir}$.

We choose to start the simulations when the center of the satellite is at a
distance of $4 \, \re$ of the host, a distance considerably smaller than
the virial radius $R_{\rm vir}$ of the host.  This starting point has the
advantage that we are still in the regime in which the stellar bulges have
not yet interacted strongly, yet it avoids the often-times long inspiral
from $R_{\rm vir}$ to $4\, \re$ that can be reasonably handled
analytically.  To map the orbital parameters from $R_{\rm vir}$ to $4\,
\re$, we have performed a series of semi-analytic calculations in which we
integrate numerically a satellite's orbit in the background potential of a
host, taking into account the processes of dynamical friction (causing
orbital decay) and tidal stripping (removing mass from outside of the Roche
limit).  The initial orbital parameters at $R_{\rm vir}$ are chosen to be
typical of those measured from cosmological simulations \citep{benson2005,
  khochfar2006}: we vary around a ``most probable'' orbit, which we take to
have $v_r \approx \vcn$ and $v_t \approx 0.7 \, \vcn$. 
For deceleration due to
dynamical friction, we use the standard formula
\begin{equation}
  \label{eq:df}
  \frac{d}{dt}\vec{v}_{\mathrm{orb}} = -4 \pi G^2 \ln \Lambda \, \msat
  \, \rho_{\rm host}(<\vorb) \, \frac{\vec{v}_{\mathrm{orb}}}{\vorb^3} \,,
\end{equation}
where $\rho_{\rm host}(<\vorb)$ is the density of host halo particles with
velocities less than the bulk velocity of the satellite and $\ln \Lambda$
is the usual Coulomb logarithm \citep{binney1987}.  [Following the results
of \citet{taylor2001}, we use
$\Lambda=1+\mhost/\msat(t)$.] The drag force is thus proportional to
$\msat^2$.  Although many of the assumptions that go into deriving the
Chandrasekhar formula are not appropriate for simulations such as ours,
previous work has shown that this equation provides a reasonable
approximation for the orbital decay of satellites when coupled with
expressions for tidal stripping and shocking (e.g.,
\citealt{velazquez1999,taylor2001,taffoni2003}).

For tidal stripping, the relevant scale is the satellite's tidal radius
$R_{\rm tid}$, which is determined by the balance between the satellite's
self-gravity and host's tides.  At separation $r$ from the host center,
$R_{\rm tid}$ is approximately given by $\overline{\rho}_{\rm sat}(R_{\rm
  tid}) \approx \overline{\rho}_{\rm host}(r)$ and the mass outside of
$R_{\rm tid}$ tends to be stripped away on a local dynamical time.  We
model this process by integrating the satellite orbit with timesteps that
are a fraction $f$ of the instantaneous orbital period, calculating the
tidal radius at each point, and removing a fraction $f$ of the mass outside
the tidal radius at each timestep.

Besides the processes of dynamical friction and tidal stripping discussed
above, impulsive heating from fast encounters is expected to lead to an
additional effect of gravitational shocking \citep{ostriker1972,
  gnedin1999a, gnedin1999}.  This effect occurs when the tidal field of the
host on the satellite varies rapidly, e.g., when a globular cluster passes
through a galactic disk or, in our case, when a satellite galaxy approaches
the stellar bulge of the host.  After initially compressing the satellite,
the shock heats the satellite and causes it to expand, rendering it more
susceptible to disruption on further pericentric passages or by tidal
stripping.  The strength of the shock (assumed to have characteristic
timescale $\tau_{\mathrm{shock}}$) depends on the satellite's internal
structure and dynamics: stars on orbits with angular frequencies $\omega
\ll \tau_{\mathrm{shock}}^{-1}$ tend to receive a boost in energy, while
orbits with $\omega \gg \tau_{\mathrm{shock}}^{-1}$ are not strongly
affected \citep{weinberg1994}.  Since our semi-analytic calculation is to
map orbital parameters between $R_{\rm vir}$ and the first crossing of $4
\, \re$, tidal shocking should be unimportant and is not included.  Our
calculation thus amounts to a simplified version of the models of
\citet{taylor2001} or \citet{taffoni2003}.  The effect of gravitational
shocking, however, is seen in our numerical simulations, as discussed in
Sec 3.

The results of our semi-analytic calculations suggest that when the
satellite-to-host mass ratio is small ($\la 0.1$) and the initial orbital
velocity at $R_{\rm vir}$ is high [$\vorb \ga \vcn(R_{\rm vir})$, where
$\vcn^2(r)\equiv GM(<r)/r$], only satellites on relatively radial orbits
(with tangential-to-radial velocity ratio of $v_t/v_r\la 1$) decay quickly
enough to be relevant, i.e., reaching $4\, \re$ within half a Hubble time.
Much of the parameter space is therefore eliminated.  For the orbits that
do decay rapidly enough, we find that the satellite velocity at $4 \,\re$
relative to the local circular velocity $\vcn$ at $4\, \re$ typically
ranges from $\vorb/ \vcn=0.7$ to 1.75.  The specific angular momentum at $4
\, \re$ tends to be high: even though we consider only orbits with $v_t/v_r
\la 1$ at the virial radius, it is not uncommon for the tangential velocity
to be greater than the radial velocity at $4 \,\re$.  As a result, we
choose initial orbital energies ranging between $\vorb/\vcn=0.75$ and 1.5,
and angular momenta ranging between $v_t/v_r=0$ and 2.0 at $4\,\re$.  The
full set of initial conditions for our simulations is listed in
Table~\ref{table-ICs}.

\section{Results: fate of satellites}
\label{sec:fate}

In this section we investigate how the survival of a merging satellite
depends on its orbit and density, as well as on the presence of a black
hole and dark matter halo.

\subsection{Dependence on orbital angular momentum and energy}
\label{sec:EL} 

\begin{figure*}
  \centering
  \includegraphics[scale=0.48]{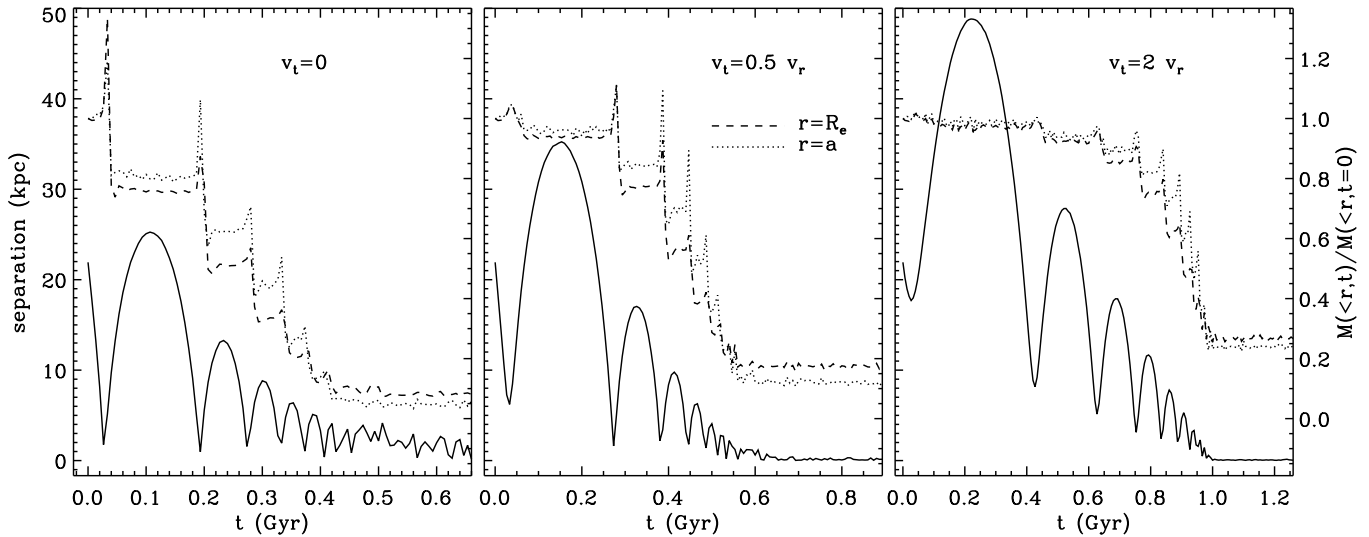}
  \caption{ Evolution of the center-of-mass separation between the
    satellite and host (solid) for three simulations that differ only
    in the initial orbital angular momentum; the ratio of
    tangential-to-radial velocities, $v_t/v_r$, increases from left to
    right.  The initial orbital energy is fixed at $\vorb/\vcn=1.5$.
    Also plotted is the evolution of the satellite's stellar mass
    $M_{\rm sat}(<r,t)$ interior to the satellite's initial effective
    radius $\re$ (dashed) and scale radius $a$ (dotted), normalized to
    the initial stellar mass within that radius.  (The scale
    corresponding to the dashed and dotted curves is given on the
    right vertical axis.)  From these plots it is apparent that most
    of the heating of the satellite bulge occurs during pericentric
    passages.  Furthermore, there is significant tidal compression
    during each pericentric passage.  Satellites on more radial orbits
    experience larger mass losses and fall to the center of the
    remnant faster.}
  \label{fig:stripDMconstE}
\end{figure*}

\begin{figure*}
  \centering
  \includegraphics[scale=0.48]{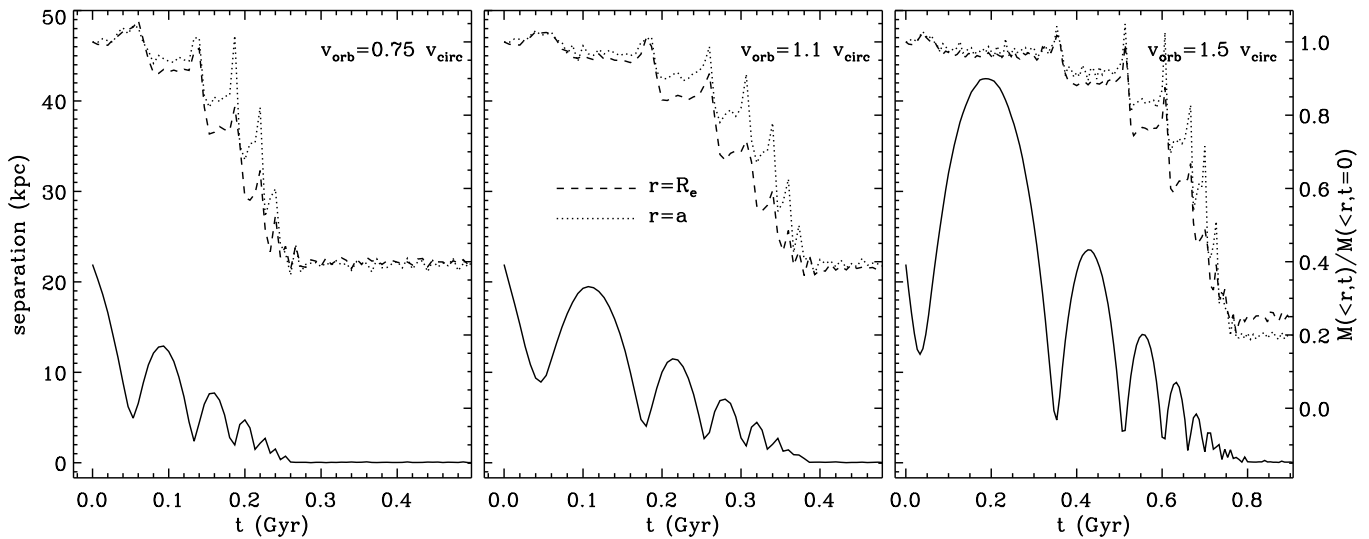}
  \caption{ Same as Fig.~\ref{fig:stripDMconstE} but for three
    simulations that differ only in the initial orbital energy.  The
    initial orbital angular momentum is fixed at $v_t=v_r$.  More
    energetic orbits lead to larger mass losses; it also takes
    significantly longer time for satellites on these orbits to decay
    to the center of the remnant.}
  \label{fig:stripDMconstL}
\end{figure*}

The three panels in Fig.~\ref{fig:stripDMconstE} compare the time evolution
of the satellite orbital decay and mass loss for three simulations that
differ only in the initial orbital angular momentum, where the initial
tangential-to-radial velocity ratio ranges from $v_t/v_r=0$ (purely radial)
to $v_t/v_r=2$ (runs H1, H2, and H4 in Table~1).  The orbital energy is
fixed at $\vorb/\vcn=1.5$.  The vertical axis plots the center-of-mass
separation between the satellite and host (solid line) and the
corresponding evolution of the mass within the satellite's initial $\re$
(dashed line) and initial scale radius $a$ (the quarter-mass radius for a
Hernquist density profile; dotted line).

Satellite mass losses are seen to follow distinct steps, with most changes
due to strong tidal shocking followed by tidal stripping as the satellite
passes through the pericenter of its orbit.  The satellite mass stays
relatively constant between pericentric approaches for all orbits.  This is
not unexpected, since both tidal stripping and gravitational shocking are
most effective at the pericenter of an orbit.  Fig.~\ref{fig:stripDMconstE}
shows that the tidal effects at the first two pericentric passages are
significantly stronger for more radial orbits, leading to faster mass loss
in the satellites.  It also shows that the orbital decay of the satellite
with low angular momentum subjects it to increasingly strong tides and
large mass loss.  The end result is that the ratio of the final to initial
satellite stellar mass within its initial effective radius $\re$ (dashed
line) is only $\approx 10\%$ for the purely radial ($v_t/v_r=0$) orbit and
increases to 18\% and 27\% for $v_t/v_r=0.5$ and 2, respectively.

A prominent feature in Fig.~\ref{fig:stripDMconstE} is the spikes in the
$M(<r,t)$ curves at each pericentric passage, which are caused by tidal
compression from the host potential accompanying a tidal shock.
\citet{dekel2003} have investigated this effect for a variety of profiles
and find that satellites in a host potential with $\gamma < 1$ experiences
compression in the radial direction (i.e. along the line separating the
centers of mass of the host and satellite) in addition to the ubiquitous
compression in the perpendicular direction.  For $\gamma > 1$, the tidal
interaction tends to stretch the satellite in the radial direction while
compressing it in the perpendicular direction.  The host galaxies in our
simulations have stellar bulges with $\ghost=0.5$, so there should be a net
compression of the satellite as it passes near the center of the host,
which is indeed the case in Fig.~\ref{fig:stripDMconstE}.

\begin{figure*}
\centering
\includegraphics[scale=0.48]{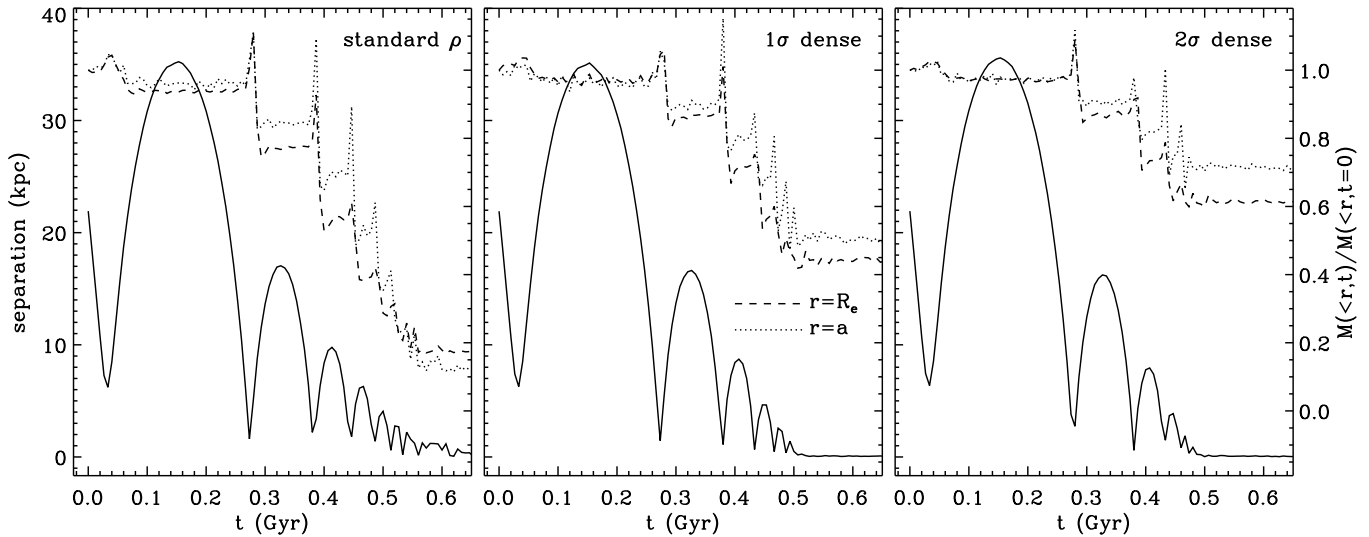}
\caption{ Similar to Figs.~\ref{fig:stripDMconstE} and
  \ref{fig:stripDMconstL} but for three simulations that differ only
  in the initial stellar density of the satellite.  Left: local
  $\re-M_{\star}$ relation from SDSS early-type galaxies.  Center:
  1$\sigma$ denser.  Right: 2$\sigma$ denser.  The runs are otherwise
  identical ($v_t/v_r= 0.5$, $\vorb/\vcn=1$, with dark matter and
  black hole).  The profiles are computed at apocenter after each
  pericentric passage.  More compact and denser satellites are
  noticeably better preserved in both the inner and outer structure.}
\label{fig:orbitDense}
\end{figure*}

Fig.~\ref{fig:stripDMconstL} is similar to Fig.~\ref{fig:stripDMconstE} but
compares results from three simulations of different orbital energies at
fixed $v_t/v_r=1$.  The corresponding runs in Table~1 are L3, M3, and H3.
The figure shows that satellites with larger initial velocity $\vorb$
(i.e. more energetic and less bound orbits) merge more slowly, experience
more pericentric passages, and suffer more severe mass loss overall.  The
amount of mass loss at {\it each} pericentric approach, however, is
relatively similar for different orbital energies, indicating that the
primary factors for determining the destructive effect of gravitational
shocks and tidal stripping on a given satellite are the number of
pericentric passages and the distance of each pericentric passage.

\subsection{Dependence on satellite density} 

Fig.~\ref{fig:orbitDense} compares the orbital decay and mass loss of the
satellite for three simulations that differ in the initial satellite
density but are otherwise identical.  The satellites in the three panels
are initially assigned the three densities shown in Fig.~\ref{fig:ICs} and
discussed in Sec.~2.1.2.  The sensitive dependence of the amount of
satellite disruption on its initial density is clearly seen: the denser the
satellite, the greater its degree of survival.  The dashed lines show the
evolution of the mass within the satellite's initial $\re$ relative to the
initial mass within the same radius.  The final value of this ratio changes
from $\sim 20\%$ for the run with the standard density to $\sim 45\%$ for
the $1\sigma$ dense run to $\sim 60\%$ for the $2\sigma$ dense run, showing
that satellite density significantly affects the satellite survival on the
scale of $\re$.  A similar effect is evident for the evolution of mass
within the satellite's scale radius $a$ (dotted lines).

The density of the satellite also influences the region from which the
satellite's mass is stripped.  The standard density satellite ends up
losing fractionally more mass within $a$ than within $\re$, while the 1 and
2$\sigma$ dense satellite lose fractionally more mass within $\re$ than
$a$.  In other words, the central region of the standard density satellite
is more affected than the half-mass radius, presumably because of the black
hole's influence.  The compact satellites are more resilient and their
outer regions are therefore relatively easier to strip than the inner
portion.

\begin{figure*}
  \centering
  \includegraphics[scale=0.5]{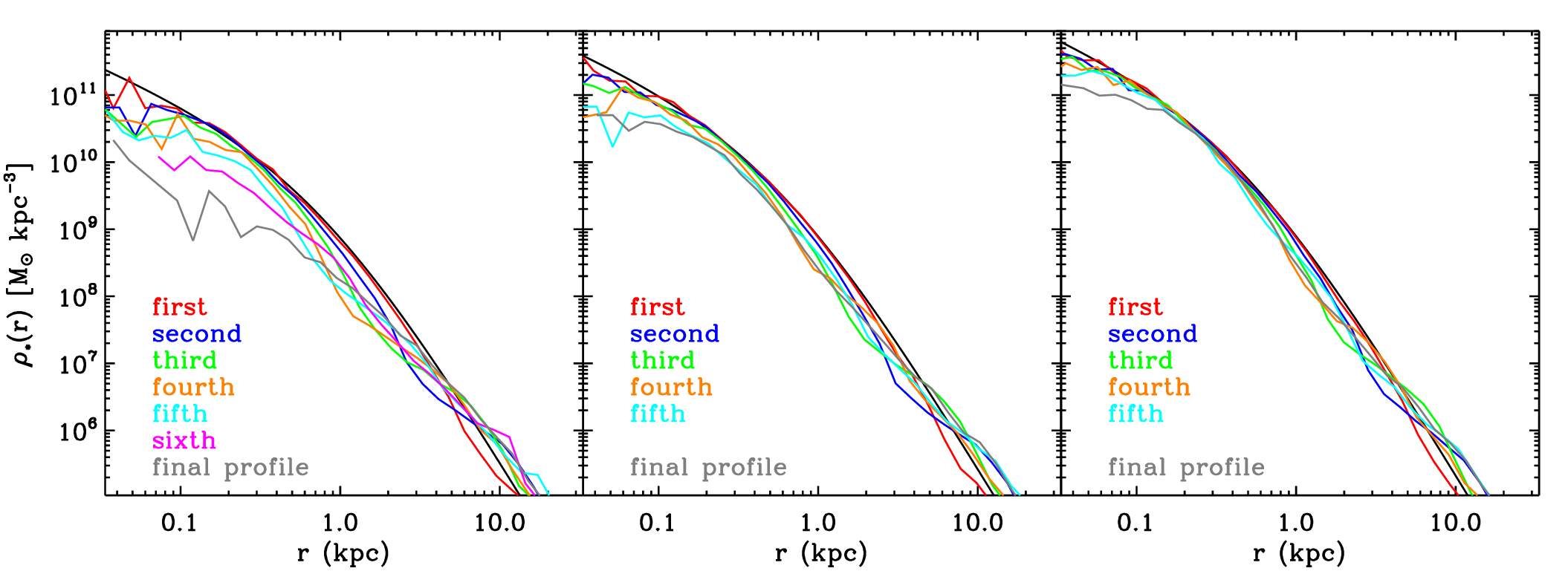}
  \includegraphics[scale=0.5]{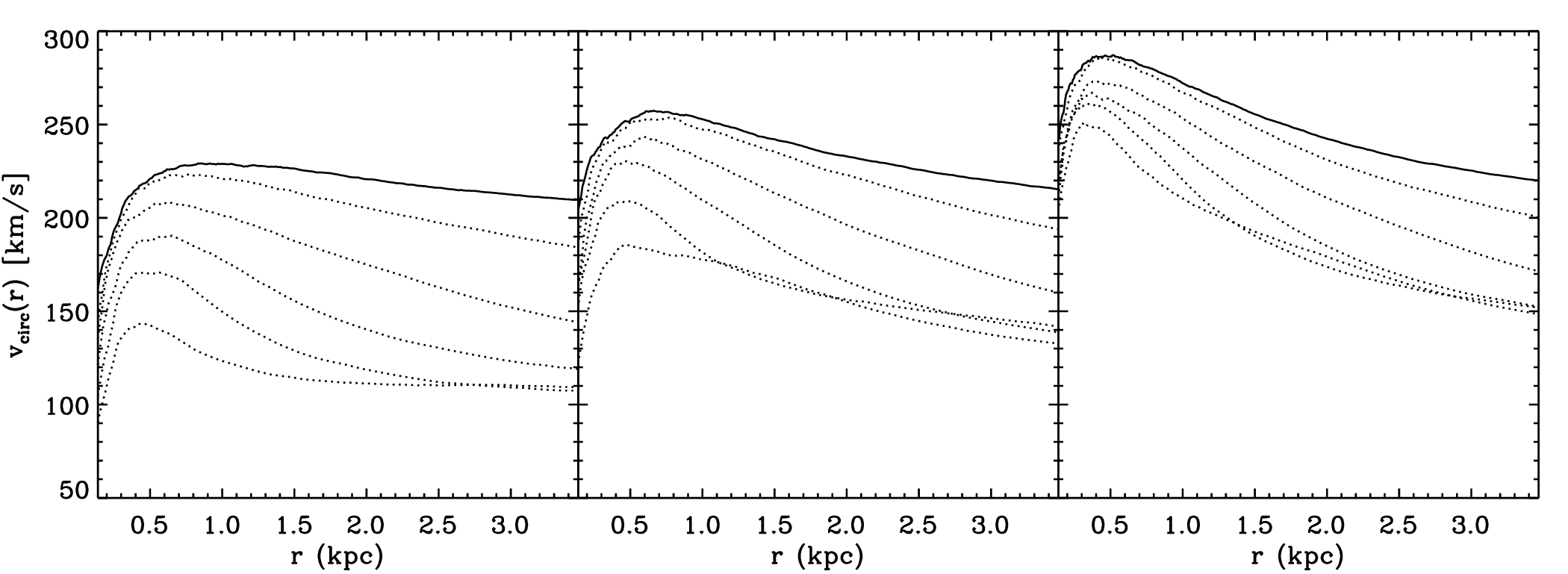}
  \caption{Evolution of the satellite's stellar density profile (top;
    logarithmic scale) and circular velocity profile (bottom; linear
    scale) for the same simulations shown in
    Fig.~\ref{fig:orbitDense}.  Left: normal (SDSS) satellite.
    Center: $1\sigma$ dense satellite.  Right: $2 \sigma$ dense
    satellite.  The profiles are computed at apocenter after each
    pericentric passage.  Denser satellites are noticeably better
    preserved, both in their inner and outer structure.  We do not
    plot the final circular velocity profile for the satellite because
    it is not meaningful when the satellite has merged with the
    host. Note the different radial scale on the upper and lower
    plots.}
  \label{fig:rhoMulti}
\end{figure*}

For a closer look at the changes in the internal structures of the
satellites as they sink towards the center of the host galaxy, we show in
Fig.~\ref{fig:rhoMulti} the satellite's stellar density profiles $\rho_{\rm
  sat}(r)$ (upper) and circular velocity profiles $v_{\rm circ}(r)$ (lower)
after each pericentric passage for the same runs as in
Fig.~\ref{fig:orbitDense}.  The profiles are computed at apocenter after
the correspondingly labeled pericentric passage, e.g., the curves labeled
``first'' correspond to profiles at apocenter after the first pericentric
pass.  The circular velocity profiles clearly illustrate that the tidal
effects during the first pericentric passage are confined to a satellite's
outer regions, while the later passages are responsible for the inner mass
loss.  The satellite with the standard density is stripped progressively
more with each pericentric passage, resulting in a final central density
that is reduced by about two orders of magnitude and a maximum circular
velocity $\vmax$ that is reduced by 40\% relative to the initial condition.
The $1\sigma$ dense satellite is much less affected in density (although
$\vmax$ is reduced by 30\%), while the most compact ($2\sigma$ denser)
satellite is stripped the least.

The effect of a black hole in the host on the satellite can also be seen
clearly in the two upper-right panels of Fig.~\ref{fig:rhoMulti}: while the
stellar density profiles of the two satellites are well preserved at
$\sim 0.5$ kpc, both profiles flatten on smaller scales due to
interactions with the black hole.  The typical radial scale on which we
might expect the effects of the black hole to appear is given by an
equation analogous to $M_\star(<\rbh)=2\, M_{\rm BH}$ in Sec.~2.1.3, with the
stellar mass referring to that of the satellite.  This gives $\rbh=0.25\,
a$ for the satellites we are considering, i.e. $\sim 0.15$ and 0.1 kpc for
the $1\sigma$ and $2\sigma$ dense satellites, respectively, which indeed
corresponds to the observed scale of the change in the density profile.  We
discuss the effects of black holes further in the next subsection.

\subsection{Dependence on presence of central black holes}

\begin{figure*}
  \centering
  \includegraphics[scale=0.6]{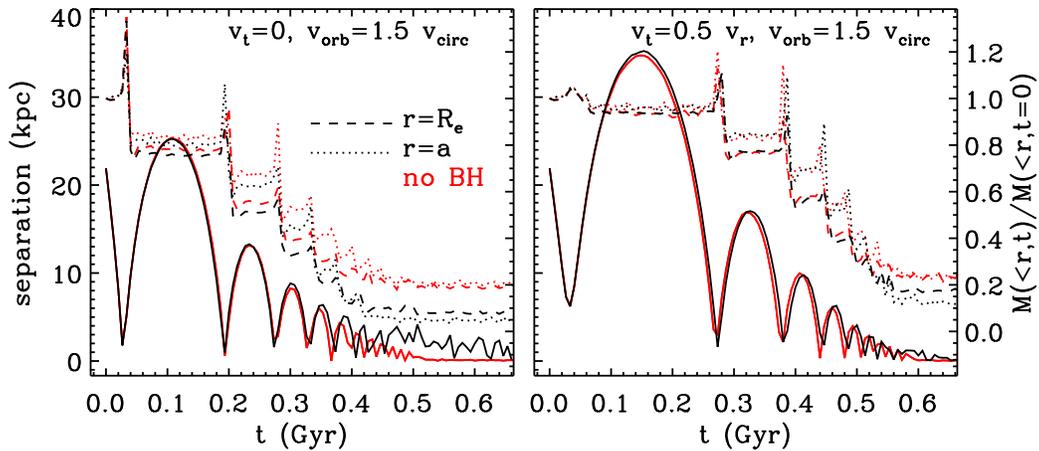}
  \caption{Comparison of the orbital decay and stellar mass loss of a
    satellite as it is accreted onto a host galaxy with (black) and without
    (red) a central black hole.  Two initial orbits are shown: radial (left
    panel) and $v_t/v_r=0.5$ (right panel).  Black holes in host galaxies
    enhance the disruption efficiency for the satellite.}
  \label{fig:stripBHnoBH}
\end{figure*}

We have performed simulations without central black holes to quantify the
effects of black holes in the host galaxies on the dynamics and disruption
of infalling satellite galaxies.  Fig.~\ref{fig:stripBHnoBH} compares
results from runs with and without black holes for two orbits (left for
radial orbit; right for $v_t/v_r=0.5$; both with orbital energy
$\vorb=1.5\vcn$).  The black hole has a non-negligible effect in the early
evolution for the radial orbit because the satellite is brought to the
host's black hole immediately.  Both the central (dotted curve) and outer
(dashed curve) regions are somewhat more disrupted in the case where the
host has a black hole.  The differences between the two runs become more
pronounced as the merger progresses: with successive pericentric passages
the black hole exerts a stronger effect, widening the gap between the black
and red curves on the left.  By the time the merger is complete, the
satellite stellar mass within its $\re$ in the black hole run is only about
half of that in the run without a black hole.  The difference is even more
pronounced at smaller radii; in fact, the black hole leads to a greater
evacuation of mass within $a$ than within $\re$, while the reverse is true
for the run without a black hole.

The runs with $v_t/v_r=0.5$ (right panel) evolve differently.  The
satellite does not pass near the black hole on the first few pericentric
passages, so the hole does not provide any shocking or stripping and there
is no difference between the runs with and without a black hole.  Once the
orbit has decayed enough to bring the satellite close to the center of the
host bulge, however, the black hole starts to exert its effect.  The final
result is again that the satellite is more destroyed when the host has a
black hole, although the effect is not as strong as for the radial
collision.

\begin{figure}
  \includegraphics[scale=0.45]{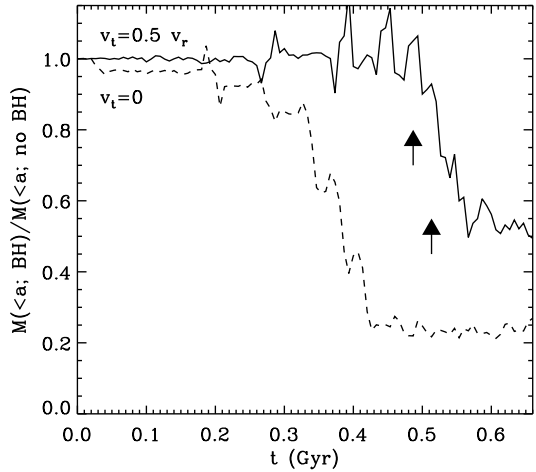}
  \caption{Evolution of the ratio of the satellite's stellar mass
    interior to the initial scale radius $a_{\mathrm{sat}}$ in runs
    with a black hole to the same quantity in runs without black
    holes.  Two orbits are shown.  Arrows mark first and second
    crossings of the black hole's sphere of influence $\rbh$ in the
    run with angular momentum.  The ratio is seen to decrease
    noticeably between the crossings due to the tidal mass stripping
    by the black hole.  The radial merger experiences crossings of
    $\rbh$ at every pericentric passage, but the first 3 encounters
    are at high velocity and thus do not lead to significant tidal
    stripping.}
  \label{fig:bh_mass}
\end{figure}

To quantify the amount of satellite destruction due to the host's black
hole versus other tidal processes, we plot as a function of time in
Fig.~\ref{fig:bh_mass} the ratio of the satellite's stellar mass within its
initial scale radius $a$, $M_{\star,{\rm sat}}(<a,t)$, for the runs with a
black hole relative to the same quantity in the runs without a black hole.
For the radial merger (dashed line), this ratio decreases rapidly to about
25\% after 0.4 Gyr due to the strong tidal effects exerted by the black
hole on the accreting satellite.  For the non-radial merger (solid line),
this ratio stays at essentially unity\footnote{There are oscillatory
  features from $t \approx 0.3$ Gyr onward, but these simply reflect the
  slightly different orbital evolution of the runs with and without black
  holes; see the red vs. black solid curves in fig.~\ref{fig:stripBHnoBH}}
until the satellite's orbit has decayed enough to cross the hole's sphere
of influence ($\rbh\sim 300$ pc) at $\approx 0.5$ Gyr.  The first two
crossings of $\rbh$ (marked by arrows) correspond quite well to the initial
decrease in $M_{\star,{\rm sat}}(<a; {\rm BH})$ relative to $M_{\star,{\rm sat}}(<a;
\mathrm{no} \, {\rm BH})$.  By the time the remnant has relaxed the run with a
black hole has $\sim 50\%$ less mass within $a_{sat}$ relative to the run
without a black hole.  This is less of a dramatic reduction than in the
radial merger because the black hole interacts fewer times with the center
of the satellite in the run with angular momentum.

We have also considered the case where both the host and the satellite have
central supermassive black holes.  In this case, each black hole mass is
set to $2\times 10^{-3}$ of its respective stellar bulge mass (see runs T5
\& T6 in Table~1).  The dense satellite (with satellite-to-host mass ratio
of 1:10) is used for this run; if there is additional disruption of the
satellite due to the presence of a black hole at its center, it will show
up better in this case than in the lower-density one.  We find that there
is no substantive difference in the density profiles of either the host or
satellite between the runs with one and two black holes.  Our simulations,
however, cannot follow the evolution of the system to the point at which
three-body interactions among a supermassive black hole binary and stars
start to scour out a core.  The separation of the black holes at the hard
binary stage is $r_{\rm hard}=G\mu/(4\sigma^2)\approx 1$ pc for the
parameters of our simulations; the highest force accuracy we use is
$\epsilon=3$ pc (run T6 in Table~1), and even this scale can be
significantly affected by two-body relaxation.  Recent simulations by
\citet{merritt2006a} have studied the evolution of black hole binaries
embedded in stellar spheroids with $\gamma$ profiles for a variety of black
hole mass ratios and values of $\gamma$.  The effect of a 1:10 binary
($q=0.1$) on a $\gamma=0.5$ profile is not very pronounced (his Fig.~5),
implying that the surface brightness profiles presented here would not be
strongly affected by core scouring.

\subsection{Comparison to runs without dark matter}

\begin{figure*}
  \centering
  \includegraphics[scale=0.6]{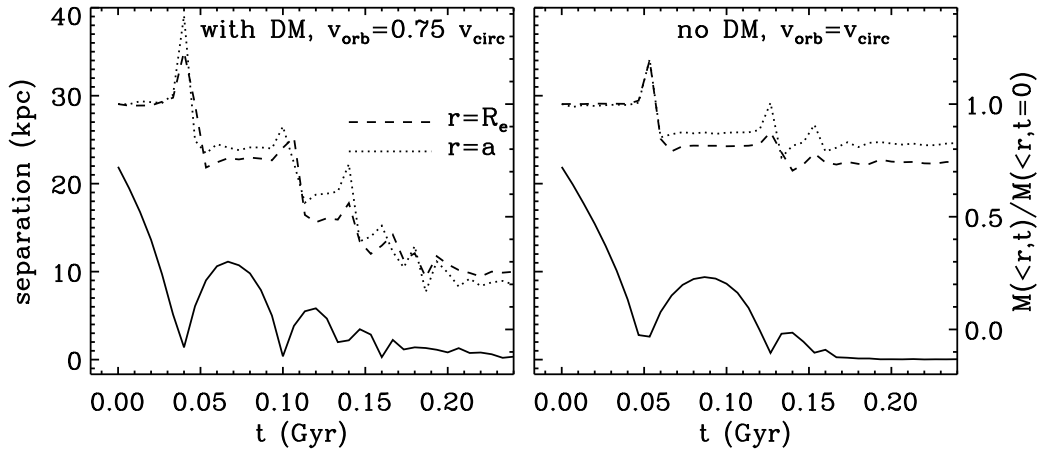}
  \caption{Comparison of radial satellite accretion with (left) and
    without (right) dark matter halos.  The run including dark matter
    (in both satellite and host) has $\vorb=0.75\,\vcn$ while the run
    without dark matter has $\vorb=\vcn$.  Based purely on the results
    presented to this point, we expect the run with the higher ratio
    of $\vorb/\vcn$ to be more disrupted.  The reverse is true,
    however, because the dark matter significantly deepens the host
    galaxy's gravitational potential, leading to more pericentric
    passages for the satellite with dark matter and hence more severe
    mass losses.}
  \label{fig:stripDMnoDM}
\end{figure*}

Prior numerical studies on satellite accretion and their survival are
focused on either purely dark matter subhalos (e.g., \citealt{hayashi2003,
  kravtsov2004, kazantzidis2004b}) or purely stellar systems (e.g., the
pioneering work of \citealt{white1983}; \citealt{balcells1990,
  weinberg1997}; black holes are included in
\citealt{holley-bockelmann2000} and \citealt{merritt2001}).  In this
subsection we compare the results of simulations ignoring the effects of
dark matter to our full simulations (using stars, dark matter, and black
holes) in order to quantify any potential differences.  In general,
simulations with only one primary component (stars or dark matter) are less
expensive to perform than simulations with both components at a fixed force
and mass resolution, so it is useful to assess whether including the
effects of dark matter is necessary for understanding satellite evolution.

As outlined in Sec.~\ref{subsec:params}, there is in fact good cause for
thinking that including both dark matter and stellar components will lead
to non-trivial differences from simulations with only dark matter halos or
stellar bulges.  The results of the previous sections indicate that one
important parameter in determining whether or not a satellite is strongly
disrupted is the number of pericentric passages it experiences before
merging with the host bulge.  Varying the depth of the host potential well
is likely to change the number of orbits a satellite undergoes.  Since
orbital deceleration is governed by dynamical friction
(equation~\ref{eq:df}), which is proportional to $M_{\rm sat}(t)/M_{\rm
  host}(t)$, and tidal stripping is likely to reduce $M_{\rm sat}$ by
removing dark matter from outside the tidal radius, we expect the
simulations with dark matter to have a slower orbital evolution and to go
through more pericentric passages before merging than the equivalent runs
without dark matter.

To quantify the effect of dark matter, we compare in
Fig.~\ref{fig:stripDMnoDM} the results from a 3-component run (with dark
matter, stars, and black hole) and a 2-component run without the dark
matter halo (run L1 and T1 in Table~1).  (The host and satellite either
both have dark matter halos or both without.)  The satellite for both runs
is on a radial orbit with initial orbital velocity $\vorb=0.75 \, \vcn$ (at
$4 \re$).  Note that the circular velocity $\vcn$ is \emph{not} equal in
the two cases, as the dark matter contributes to the circular velocity.
Fig.~\ref{fig:stripDMnoDM} shows that the satellite in the run with dark
matter (left panel) experiences more pericentric passages before the final
merger than the run without dark matter (right panel).  Even though the
satellites' stellar masses interior to $\re$ and $a$ are quite similar
between the two runs after the first pericentric passage, the additional
$\sim 3$ passages in the run with dark matter cause extra disruption.  As a
result, only 20-30\% of the satellite's stellar mass remains within its
initial $\re$ and $a$ after 0.2 Gyr in the 3-component run (dashed and
dotted line in left panel) compared with the 70-80\% in the run that
ignored dark matter.

\subsection{Other factors: mass ratio, inner density profile}
\label{sec:mass_gamma}

We have tested a number of other parameters that can potentially affect the
outcome of our satellite accretion simulations.  Primary among these are
the satellite-to-host mass ratio and the inner logarithmic slope $\gsat$
of the satellite's density profile.

All the results presented thus far have assumed a satellite-to-host mass
ratio of 1:10.  We have performed a two-component test simulation (run T2
in Table~1) with a lower mass ratio of 1:33 and found qualitatively similar
results.  These lighter satellites lose their orbital energy on a longer
time scale, however: it takes a factor of $\sim$1.5-2.5 longer for their
stellar bulges to merge with and have observational impact on the host's
bulge.  In a cosmological setting, we therefore expect the satellites with
$\la 5$\% of the host mass to have less direct dynamical effect on the host
than the more massive satellites despite their higher number density.

We have also performed several simulations (runs labeled ``s'' in Table~1)
using a steeper $\rho\propto r^{-1.5}$ cusp for the initial stellar density
of the satellite.  As expected, these satellites suffer less tidal
disruption than the $r^{-1}$ satellites studied thus far, but the net
result is somewhat degenerate with decreasing $\re$ of the satellite while
fixing the $r^{-1}$ cusp.  For example, our standard density satellite
($M_{\star}=3.3 \times 10^{10}\,M_{\odot}, \, a=0.76 \, \mathrm{kpc}, \,
\gsat=1$) looks quite similar to a satellite with the same stellar mass but
$\gsat=1.5$ and $a=1.35$ kpc: the densities differ by a maximum of 30\%
from 30 pc (our standard force softening) to 5 kpc.  We include analyses of
the runs with steeper satellite stellar density in the sections that follow
below.

\section{Observable consequences of satellite accretion}
\label{sec:observables}

\subsection{Surface brightness profiles}

\begin{figure*}
  \centering
   \includegraphics[scale=0.48]{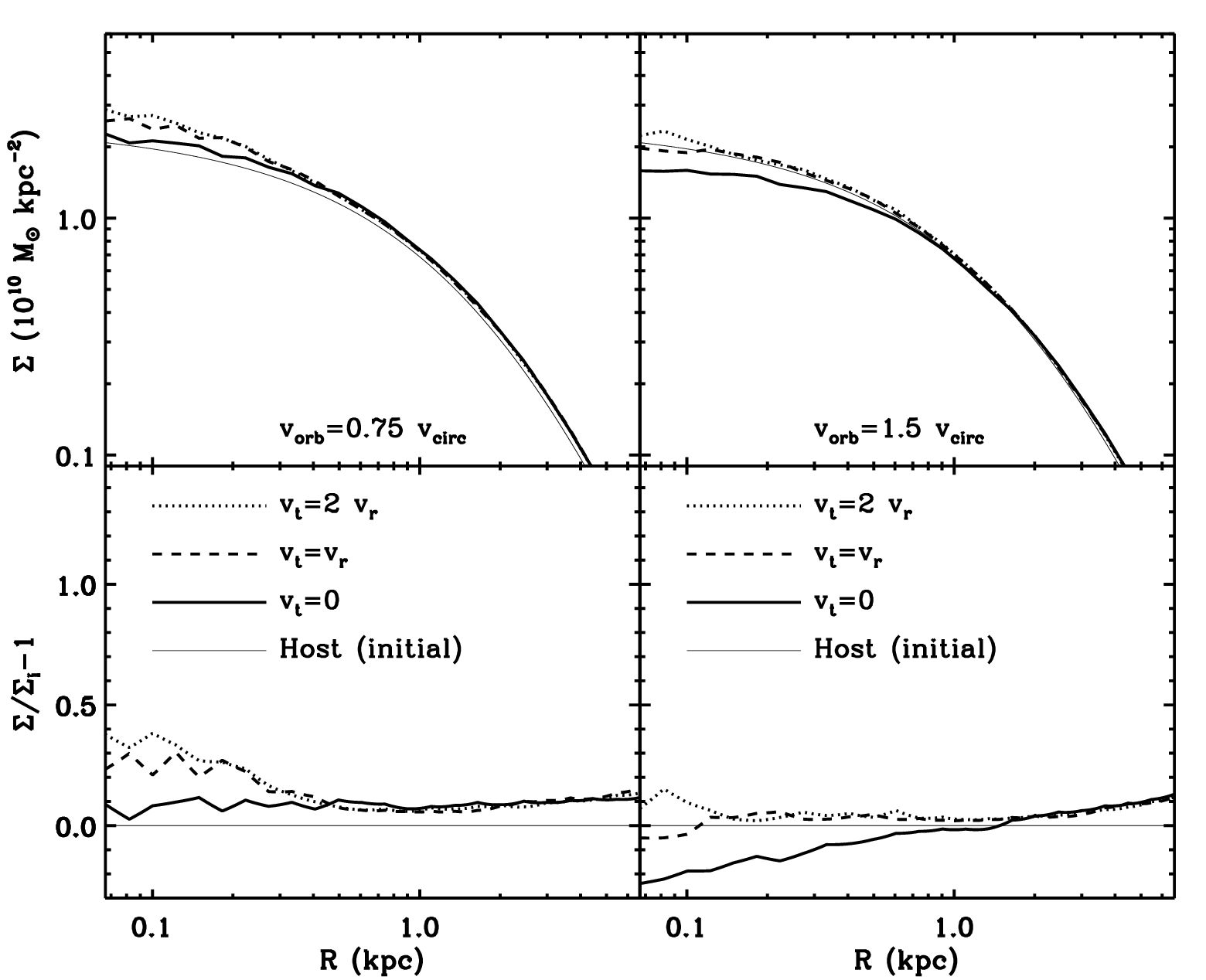}
   \includegraphics[scale=0.48]{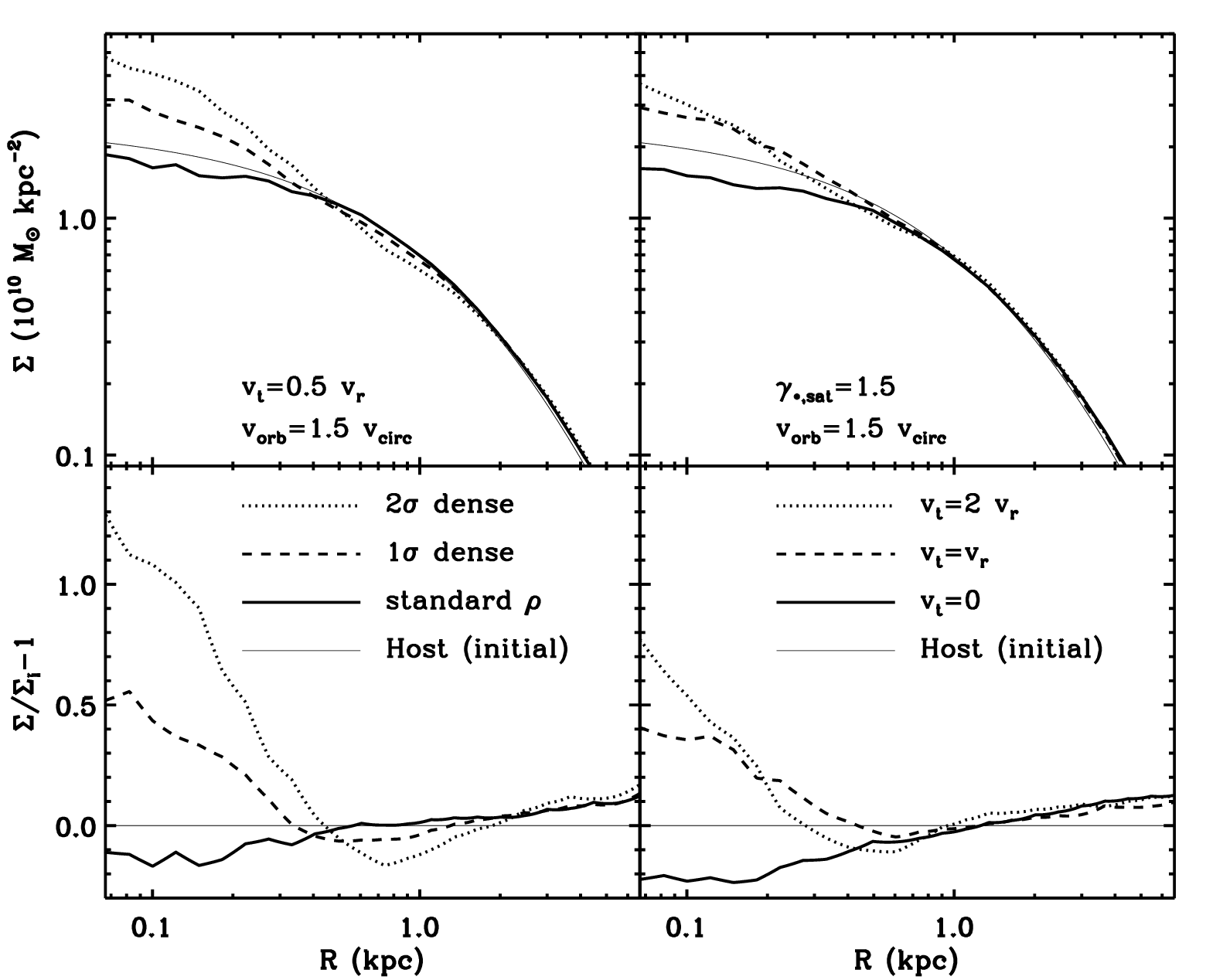}
   \caption{ Upper of each set of panels: Surface brightness profiles
     $\Sigma(R)$ of the host galaxy initially (thin solid curves) and
     after accretion of a satellite (computed from all stellar
     particles) for selected three-component runs.  Lower of each set
     of panels: Same as top panels except it shows the corresponding
     fractional change between the initial and final $\Sigma(R)$ to
     accentuate the evolution of the surface brightness.  Upper left:
     runs with varying orbital angular momentum at fixed orbital
     energy of $\vorb/\vcn=0.75$ and $\gsat=1.0$.  The final central
     surface brightness is larger than the initial for all three runs
     but the amount of increase decreases with decreasing angular
     momentum.  Upper right: runs with varying orbital angular
     momentum at fixed orbital energy of $\vorb/\vcn=1.50$ and
     $\gsat=1.0$.  The increased orbital velocity leads to less
     deposition of satellite mass at the center of the remnant than in
     the $\vorb/\vcn=0.75$ case, and the radial collision results in a
     noticeably reduced inner central surface brightness.  Lower left:
     runs with fixed orbital angular momentum ($v_t/v_r=0.5$) and
     fixed orbital energy ($\vorb/\vcn=1.5$) with $\gsat=1.0$ and
     varying satellite stellar density (i.e. varying $\re$).  The
     standard density satellite results in a reduced central surface
     brightness while the more compact satellites have dense centers
     that survive to the center of the host bulge, adding a large
     amount of mass.  Lower right: runs with varying orbital angular
     momentum at fixed orbital energy of $\vorb/\vcn=1.5$ and
     $\gsat=1.5$.  The final central surface brightness decreases for
     the radial run but increases for both runs with angular
     momentum.}
  \label{fig:sb_profiles}
\end{figure*}
  
As we have shown in Sec~\ref{sec:fate}, the fate of a sinking satellite
depends sensitively on a number of parameters including its orbital energy
and angular momentum, its mass concentration, and the presence of black
holes and dark matter halos.  Even one dense satellite galaxy that finds
its way to the center of a massive elliptical galaxy without being
significantly disrupted can lead to a central density enhancement that is
occasionally observed, so surface brightness profiles can serve as a useful
diagnostic for structure formation models.

\begin{figure}
  \centering
  \includegraphics[scale=0.5]{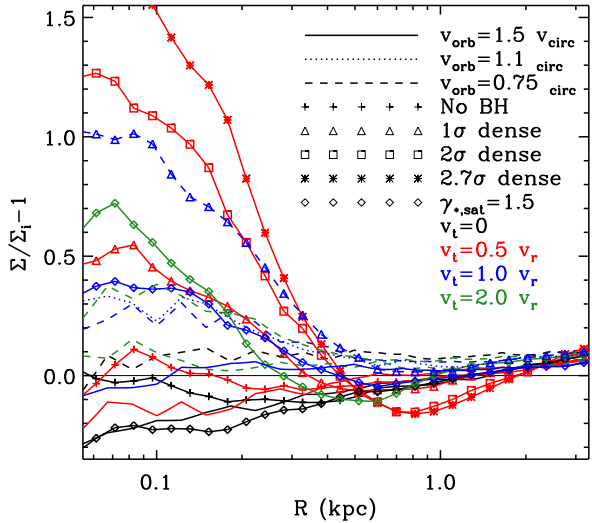}
  \caption{Fractional change in the central surface brightness profile
    of a host galaxy after accretion of a satellite from all
    3-component simulations listed in Table~1.  The profiles show a
    wide spectrum of behavior on small radial scales, ranging from
    reduction in the central surface brightness for energetic and
    eccentric orbits, to sharp increases when the accreted satellite
    is relatively compact and dense initially.  All the final surface
    brightness profiles have similar behavior at larger radii ($\ga 2$
    kpc), regardless of the central profile.}
  \label{fig:sb_all}
\end{figure}

In Fig.~\ref{fig:sb_profiles} we illustrate how a single sinking satellite
on a variety of orbits and with a range of stellar densities can modify the
shallow surface brightness profile of a massive elliptical galaxy described
in Sec~2.  Fig.~\ref{fig:sb_all} is a busier summary plot for all of our
3-component simulations listed in Table~1 (including those in
Fig.~\ref{fig:sb_profiles}).  Each curve is computed using the mean of
twenty random projections of the merger remnant.  In
Fig.~\ref{fig:sb_profiles}, the upper left panels compare a relatively
bound satellite orbit at fixed energy ($\vorb/\vcn=0.75$) with varying
angular momentum; the upper right panels show a more energetic orbit
($\vorb/\vcn=1.5$) with varying angular momentum; the lower left panels
compare runs with identical orbital parameters ($\vorb/\vcn=1.5$,
$v_t/v_r=0.5$) but varying initial satellite densities; and the lower right
panels compare energetic ($\vorb/\vcn=1.5$) mergers with steeper satellite
stellar density ($\gsat=1.5$) and
varying angular momentum.  The top panels show the absolute surface
brightness profile $\Sigma(R)$ of the stellar component in the host galaxy
initially (thin solid lines) and after satellite accretion (other lines;
computed from all stellar particles in both the host and satellite); the
bottom panels show the fractional changes between the initial and final
$\Sigma(R)$.

The general trend in the upper two sets of panels in
Fig.~\ref{fig:sb_profiles} and in Fig.~\ref{fig:sb_all} is that satellites
on more energetic or more eccentric (i.e. lower angular momentum) orbits
are more destroyed due to the stronger tidal effects.  This trend is
consistent with the results in Sec.~\ref{sec:EL} and
Figs.~\ref{fig:stripDMconstE} and \ref{fig:stripDMconstL}.  While
Figs.~\ref{fig:stripDMconstE} and \ref{fig:stripDMconstL} illustrate the
orbital decay and mass loss of the satellite, Figs.~\ref{fig:sb_profiles}
and \ref{fig:sb_all} show the net effect of the sinking satellite on the
inner stellar distribution of the host galaxy.

A particularly noteworthy feature is seen in several cases in
Fig.~\ref{fig:sb_profiles} and in Fig.~\ref{fig:sb_all}: the central
surface brightness of a core elliptical galaxy is actually \emph{reduced}
when a satellite is accreted on a relatively energetic ($\vorb \ga \vcn$)
and eccentric ($v_t \la v_r$) orbit.  Satellite accretion in this
particular orbital parameter space is likely to occur infrequently, but it
provides a potential mechanism for producing a minimum in the central
surface brightness.  Minima \emph{are} seen in $\sim 5$\% of the core
galaxies \citep{lauer2002, lauer2005}, although their physical scale tends
to be at $\la 100$ pc.  A broader search in parameter space would be needed
to determine whether the observed features can be more closely reproduced
by specific combinations of parameters in this satellite accretion
scenario.

The somewhat counter-intuitive reduction in central $\Sigma(R)$ via
satellite accretion is the result of two processes.  First, the accreting
satellite clearly has to be largely destroyed so as to bring in little
stellar mass to the host's center.  As Figs.~\ref{fig:stripDMconstE} and
\ref{fig:stripDMconstL} showed, energetic and eccentric orbits are
effective means of destroying satellites.  Second, as a satellite sinks via
dynamical friction toward the host's center, energy is transferred from the
satellite to the host bulge, leading to gravitational heating that helps
reduce the central stellar density in the host.  Such satellite-host
dynamical interaction and its effect on the bulge density and surface
brightness profiles is therefore similar to that for dark matter halo
density profiles, where dark matter subhalo accretion tends to heat the
central cusp, reducing the central density, while deposition of the
stripped mass and merging of the subhalo tends add to the central density
\citep{ma2004a}.  In both cases (satellite accretion and dark matter
subhalos) it is the competition between gravitational heating and mass
deposition that determines whether the final profile is more dense or more
diffuse than the initial profile.

The lower left panels of Fig.~\ref{fig:sb_profiles} show the strong effect
of the initial satellite density on the host's central $\Sigma(R)$.  The
satellite's orbital decay, mass loss, and the evolution of its interior
structure from the same three simulations were shown in
Figs.~\ref{fig:orbitDense} and \ref{fig:rhoMulti}.  While the satellite on
the SDSS $\re-M_\star$ relation (thick solid lines) is disrupted and leads
to a $\sim 10$\% overall reduction in $\Sigma(R)$ in the inner $\sim 500$
pc (see paragraph above), the denser satellites (dashed and dotted lines)
lead to a dramatic increase in the central surface brightness.  Similar
results are seen for satellites with $\gsat=1.5$ rather than $1.0$ with
substantial angular momentum (lower right panels). 
Accretion of dense satellite -- either with a steeper central stellar
density profile or with a smaller $\re$ -- is therefore likely to add a
dense remnant to the center of a core galaxy even when the tidal effects
from the black holes are included.  A small number of surface brightness
profiles from the galaxy sample in \citet{lauer2005} somewhat resemble the
profiles of the $1-2\sigma$ dense satellites (e.g., NGC 3384).  We suggest
that satellite accretion offer a plausible explanation for the shapes of
those profiles (see also \citealt{kormendy1984}).

\subsection{Colors}

\begin{figure*}
  \centering
  \includegraphics[scale=0.45]{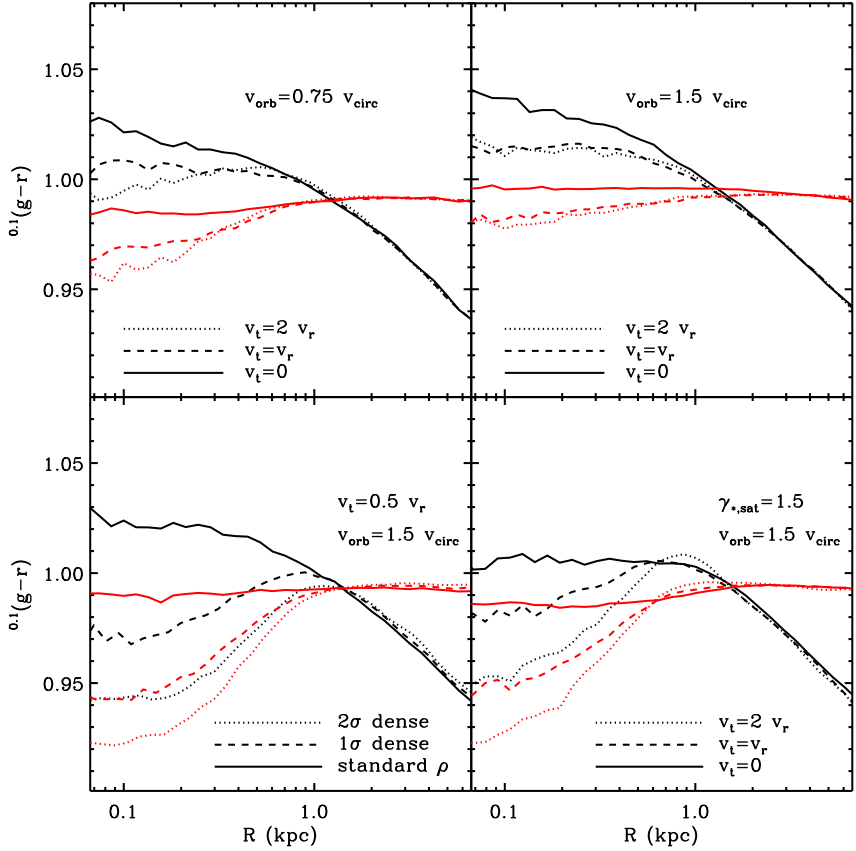}
  \caption{$^{0.1}(g-r)$ color profiles for mergers with dark matter and a
    black hole, each averaged over 20 random projections.  The host is
    assumed to have a color of 1.0 initially, while the satellite is given
    a color of 0.9 in each case.  We consider two cases: no color gradient
    in either galaxy (red curves) or a color gradient of -0.1 mag/dex in
    both satellite and host (black curves).  The panels are arranged as in
    Fig.~\ref{fig:sb_profiles}.  Upper left: $\vorb=0.75 \,\vcn$ with varying
    $v_t/v_r$.  Upper right: $\vorb=1.5 \,\vcn$ with varying $v_t/v_r$.
    Lower left:
    $\vorb=1.5\, \vcn, \, v_t/v_r=0.5$ with varying satellite density.
    Accretion of a compact satellite leads to a relatively bluer central
    remnant, while standard density satellites cause less change regardless
    of orbit.  Lower right: $\vorb=1.5 \, \vcn$ and $\gsat=1.5$ with varying
    $v_t/v_r$.}
  \label{fig:colors}
\end{figure*}

\begin{figure*}
  \centering
 \includegraphics[scale=0.45]{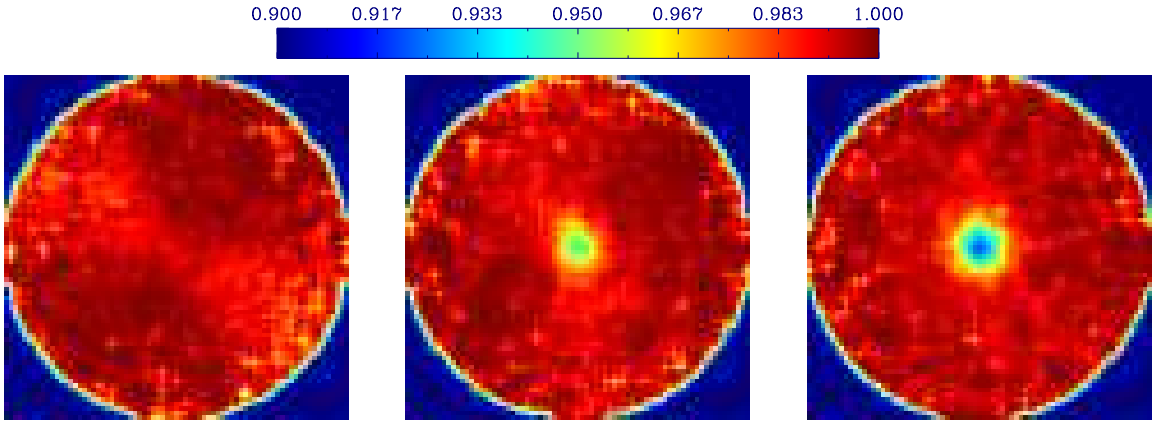}
 \caption{ Two-dimensional color map of the host bulge, viewed in the
   orbital plane, after accretion of a satellite with different initial
   densities: standard (left), $1\sigma$ denser (middle), and $2\sigma$
   denser (right).  The satellite orbit is identical for all three cases:
   $v_t/v_r=0.5, \, \vorb/\vcn=1.5$; neither galaxy is given an initial
   color gradient.  Each panel is (6.66 kpc)$^2$; each
   pixel corresponds to 0.1 kpc.  The cores of the 1 and 2$\sigma$ dense
   satellites survive to leave a visible core bluer by $\approx 0.05-0.08$
   mag.  Close inspection of the left panel shows that although the
   satellite with standard density does not survive to the center, it
   leaves faint plumes of stripped stars (light red) at very low color
   contrast ($\approx 0.01$ mag bluer).}
  \label{fig:colorMulti}
\end{figure*}

Both the average color and the color gradients in elliptical galaxies
contain vital information about their star formation and assembly
histories.  For instance, the slope and scatter along the red sequence in
the color-magnitude relation (CMR) is a useful diagnostic of how
pre-existing early-type galaxies build up larger elliptical galaxies, as
this merging would tend to average the colors and thus wash out any
correlations \citep{bower1998}.  Monolithic collapse models generally
predict steep color gradients originating from strong radial metallicity
gradients \citep{larson1974, carlberg1984}.  The case is not as clear-cut
in hierarchical models, as the effects of metallicity gradients established
by gas-rich mergers can be diluted by violent relaxation and mixing in
gas-poor mergers.  In general, the distribution of a satellite's stars
should affect the local color of a host in a way that depends on the
satellite orbit and density.  If the satellite sinks to the host's center
without substantial mass loss, for example, the resulting color on small
scales will be close to the satellite's initial color.  If, on the other
hand, the satellite is disrupted outside the host's core, the remnant's
central color will be essentially identical to that of the host initially.

We investigate the changes in galaxy colors due to satellite accretion by
assigning initial colors to the host and satellite and then computing the
remnant's color under the assumption of no change in color due to stellar
population evolution.  This method isolates the effects of mixing in the
mergers.  We assign the initial galaxies $^{0.1}(g-r)$ colors\footnote{The
  exponent 0.1 in the (g-r) color indicates that the SDSS galaxy magnitudes
  are K-corrected to z=0.1 rather than z=0.0 -- see,
  e.g. \citet{hogg2004}.} of 0.9 for the satellite and 1.0 for the host.
The observed CMR slope from SDSS in $^{0.1}(g-r)$ is -0.022 mag/dex
\citep{hogg2004}, meaning we have slightly accentuated the expected color
difference between the host and satellite given their masses.  When we
consider galaxies with initial color gradients, this color corresponds to
the color measured within the effective radius.

Fig.~\ref{fig:colors} shows the circularly-averaged color profiles for the
same sets of simulations shown in Fig.~\ref{fig:sb_profiles}; as in
Fig.~\ref{fig:sb_profiles}, each curve is computed using the mean of twenty
random projections of the remnant.  The red curves assume no initial color
gradient for either galaxy and show that the properties of the satellite
and its orbit make a significant difference in the color of the inner $\sim
1$ kpc of the remnant.  The dense satellites (lower left panel) and
$\gsat=1.5$ satellites (lower right panel) lead to a bluer center, a result
of their undisrupted cores settling to the center of the host.  The runs
with standard density and varying orbital properties show less pronounced
central differences.  For all simulations, the $^{0.1}(g-r)$ color at large
radii is about 0.99, so $\Delta$ color $\approx 0.01$ mag.

The black curves in Fig.~\ref{fig:colors} assume each galaxy initially had
a color gradient of $-0.1$ mag/dex and show how these gradients evolve as a
result of the merger.  In the runs with lower $v_{\mathrm{orb}}$ (upper
left panel), the gradient is diluted somewhat in the center but only the
run with the highest angular momentum causes a central inversion.  None of
the runs at higher $v_{\mathrm{orb}}$ and $\gsat=1.0$ (upper right panel)
have a central color inversion, but the color profile does flatten toward
the center.  The runs with higher stellar density or steeper inner stellar
density slope show a marked change in
central color at approximately $\re/5$ (as in the case without initial
gradients).  The general trend we see is to weaken the central color
gradients while affecting the outer regions less strongly.  This weakening
is in agreement with the trend reported in \citet{white1980}.  Luminous
elliptical galaxies do tend to have slightly weaker gradients in
metallicity \citep{carollo1993} and color \citep{lauer2005} than faint
ellipticals galaxies, perhaps as a result of this mixing process.

To quantify the color signatures of satellite accretion further, we have
also examined two-dimensional color maps of the hosts to look for features
that are not evident in the 1-dimensional color profiles shown in
Fig.~\ref{fig:colors}.  A set of such maps (viewed in the orbital plane) is
shown in Fig.~\ref{fig:colorMulti}, comparing the three runs with different
initial satellite densities.  The initial colors are the same as for
Fig.~\ref{fig:colors}.  The 2$\sigma$ dense satellite (right panel) sinks
to the center without significant disruption, leaving an observable core of
$\approx 0.5$ kpc in the remnant.  Even the $1\sigma$ dense satellite
(center panel) leaves an observable color signature in the core, although
it is both smaller and and less concentrated than the $2\sigma$ dense core.
By contrast, the standard density satellite is largely destroyed and leaves
no core in the remnant (left panel of Fig.~\ref{fig:colorMulti}), as was
previously seen in other ways (e.g., Figs~\ref{fig:orbitDense} and
\ref{fig:sb_profiles}).  Close inspection, however, shows faint plumes of
bluer features at low color differences ($\approx 0.01-0.02$ mag) from
stripped stars of the torn-up satellite.  The axis of the plume correlates
well with the orbit of the satellite over its final several pericentric
passages (which are quite radial).

\subsection{Additional observable properties: intra-cluster light and
  kinematic structure}

Many massive galaxies at or near the centers of clusters are cD galaxies
with extended stellar envelopes, often extending out to hundreds of kpc.
Satellite accretion onto cluster halos (and subsequent
satellite disruption) contributes, perhaps significantly, to this
intracluster light (ICL; e.g., \citealt{merritt1984, moore1996, gregg1998,
  gnedin2003, zibetti2005, mihos2005}).  Our
initial setup for the galaxies is biased against producing ICL in the sense 
that we truncate the stellar bulge, removing the stars that are at large
radii (and thus are least bound).  This truncation eliminates the stars
that would be most easily removed from the satellite and deposited at large
radii relative to the center of the remnant.  Nevertheless, it is not
inconceivable that a fraction of stars in our mergers is deposited out to
large radii, contributing to an ICL component.

The remnants in all of our simulations show excess stellar mass at large
radii ($>3 \re$ for the initial host profile), as is visible in the lower
panels of Fig.~\ref{fig:sb_profiles} and Fig.~\ref{fig:sb_all}.  Although
the excess is relatively small -- of order 15-25\% -- it shows that the
star particles do get deposited significantly outside of the host's
effective radius, indicating that tidal stripping of merging satellites can
contribute to ICL.  For cases with $\vorb=1.5\, \vcn$, the stellar mass
that belonged to the satellite tends to be distributed with a larger
characteristic radius than the mass that belonged to the host.  Simulations
of repeated satellite accretion in a cluster environment would likely yield
a significant ICL component originating from the tidal stripping of
satellites.  Analyses of cosmological simulations have shown this process
to yield some ICL \citep{sommer-larsen2005, willman2004, rudick2006};
future higher-resolution simulations should shed more light on the origin
of the ICL.

Many elliptical galaxies are observed to have kinematically distinct
substructure at their centers (kinematically decoupled cores, KDCs); often
these cores counter-rotate relative to the galaxy \citep{de-zeeuw1991}.  A
recent survey by the SAURON team of 48 nearby ($cz < 3000 \, {\rm km}\,
{\rm s}^{-1}$) E/S0 galaxies revealed a wide range of kinematic
substructure, from $\sim 100$ pc young cores in fast rotators to kpc-scale,
old KDCs in slow rotators \citep{mcdermid2006}.  We have tested for such
substructure by making velocity maps of our remnants.  None of them exhibit
a KDC; the velocity fields in all cases are consistent with zero net
rotation and no kinematic subcomponents.  Intriguingly, the three most
luminous galaxies in the SAURON sample, including M87, do \emph{not} have
KDCs.  Since this type of galaxy is similar to our initial host, our
simulations are in this sense in good agreement with the SAURON
observations.

A broader question to be explored in future work is the origin and
fragility of KDCs in slow-rotating galaxies.  Gas physics could play a role
in creating KDCs: \citet{hernquist1991}, \citet{jesseit2006}, and
\citet{cox2006} have demonstrated that simulations of binary disk galaxy
mergers with gas can lead to kinematic substructure in the remnants.  In
dissipationless simulations, \citet{balcells1990} have shown that a dense
satellite sinking to the center of a more diffuse host can create a
counter-rotating core {\it if} a large initial rotation of
$V/\sigma_0=0.75$ for the satellite and 0.4 for the host are used and the
merging orbits are retrograde.  It is somewhat difficult to discern from
the SAURON results what fraction of slow-rotators should have KDCs given
the sample size (12 slow-rotators).  If their results are representative,
however, then it is possible that any KDC would be erased if massive
ellipticals are built by multiple generations of merging.

One satellite accretion event with mass ratio 1:10 does not strongly affect
the radius or velocity dispersions of the remnant nor does it increase
$M_\star$ substantially, meaning scaling relations such as the fundamental
plane or its projections do not change notably.  Remnants of major mergers
(mass ratio 1:1-1:3), on the other hand, can have very different properties
depending on energy or angular momentum of the merger orbit
\citep{boylan-kolchin2006}.
Checking the evolution of sizes, stellar masses, and velocity
dispersion over a sequence of merger events would therefore be a good test
of the evolution of scaling laws such as the Faber-Jackson and $\mbh-\sige$
relations.

\section{Conclusions and Discussion}

In this paper we have used dissipationless simulations to examine in detail
and over a wide range of parameter space the case-by-case outcome of a
single satellite elliptical galaxy accreting onto a core elliptical galaxy.
We have taken special care to explore cosmologically relevant orbital
parameters and to set up realistic galaxy models that include all three
relevant dynamical components: dark matter halos, stellar bulges, and
central massive black holes.  The main results in this paper are as
follows:

(i) Satellites on more energetic or more eccentric (lower angular momentum)
orbits are destroyed more efficiently due to the stronger tidal effects from
the host spheroid and the central black hole (see Figs.~2, 3, 9-12).

(ii) Denser satellite galaxies -- those with either a smaller $\re$ or a
steeper inner slope -- survive significantly better (see
Figs.~\ref{fig:orbitDense}, 5, 9-12).  For instance, the standard density
satellite in Fig.~\ref{fig:orbitDense} loses more mass by its 3rd
pericentric passage than the total mass loss in the $2\sigma$ dense
satellite during its entire orbital evolution.

(iii) A supermassive black hole at the center of the host galaxy
significantly increases the efficiency with which an accreting satellite is
destroyed (see Figs.~6, 7, 10).  Without a central black hole, luminous
elliptical galaxies would have more difficulty in protecting their shallow
inner profile (however formed initially) from steepening due to mass
deposit through satellite accretion, in agreement with previous
2-component numerical work \citep{holley-bockelmann2000, merritt2001}.  

(iv) Ignoring dark matter halos in the galaxy models causes significant
differences primarily because dark matter halos lead to higher orbital
velocities for a fixed ratio of $\vorb/\vcn$ (see Fig.~8).  Satellites with
increased orbital velocity complete more orbits before merging with the
host, resulting in greater destruction of the satellite galaxy.

(v) Our overall conclusion is that the accretion of a single satellite
elliptical galaxy (1/30 to 1/10 mass of the host) can result in a broad
range of changes, in both signs, in the surface brightness profile
$\Sigma(R)$ and color of the central part of an elliptical galaxy, as
summarized in Figs.~\ref{fig:sb_all}-12.

More specifically, Figs.~\ref{fig:sb_all}-12 show that when a black hole is
included in the host galaxy 
and when the satellite and host are both
on the SDSS stellar size-mass relation, the fractional change in
$\Sigma(R)$ in the inner $\sim 500$ pc of the host galaxy upon satellite
accretion ranges from $\sim -20\%$ to $\sim +40\%$ When more compact
satellites are considered in accordance to the large scatter about the SDSS
size-mass relation, the inner $\Sigma(R)$ of the host galaxy can increase
by more than 100\%.  Since accretion of a single satellite adds little mass
and luminosity to the host, we suggest that a ``core'' galaxy at $M_v\sim
-22$ can potentially transition into a galaxy with power-law
characteristics at a comparable luminosity by accretion of a relatively
compact and massive ($\sim$ 1:10) satellite.  A core-in-core structure
would provide particularly strong evidence for this process (see also
\citealt{kormendy1984}).

The strong dependence of satellite mass loss on the satellite's density
relative to the host and the inclusion of black holes has broader
implications, for example, in interpreting the connection between dark
matter subhalos and galaxies in cosmological simulations that model only
dark matter.  Dark matter subhalos in these simulations are often stripped
to the point of not being distinguishable from the host dark matter halo.
In semi-analytic models of galaxy formation (e.g. \citealt{gao2004a,
  wang2006}), these subhalos are tracked as ``galaxies'' by either
following the most bound particle of the subhalo or extrapolating their
positions and velocities using information from previous timesteps.  The
results from our three-component (dark matter+stars+black hole)
non-cosmological simulations suggest that the properties of the observable
stellar component in the simulations are very sensitive to their assumed
initial properties, and in particular, to the initial stellar density
assigned to the galaxies and the inclusion of black holes in the host
galaxies.  Future work that incorporates the results presented here into a
full cosmological context should be valuable.

\acknowledgements

We thank E. Quataert and A. Kravtsov for useful discussions.  C--PM is
supported in part by NSF grant AST 0407351 and NASA grant NAG5-12173.  This
work used resources from NERSC, which is supported by the US Department of
Energy.

\appendix
\section{Properties of two-component $\gamma$ models}
\label{sec:append1}
\begin{figure}
  \centering
  \includegraphics[scale=0.55]{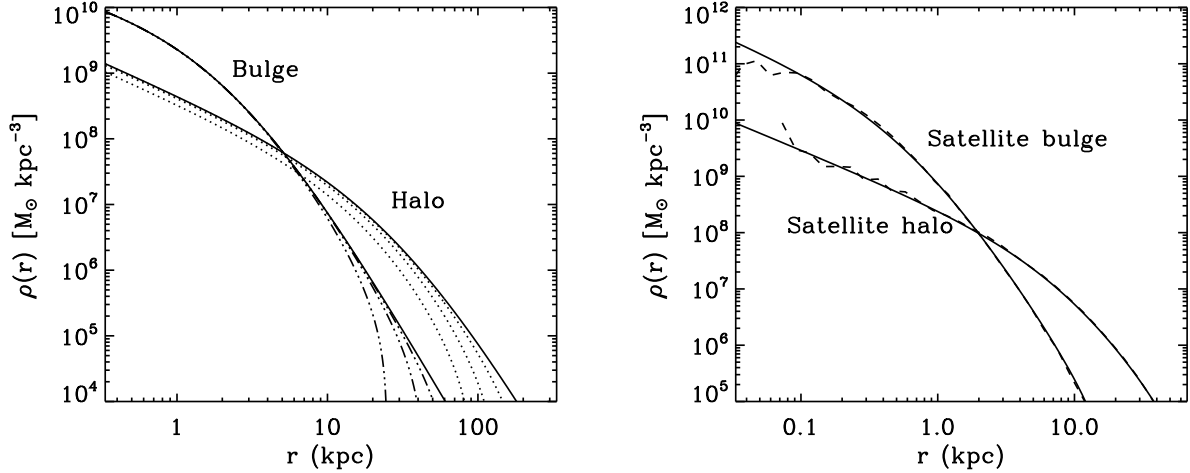}
  \caption{\emph{Left:} Density profiles of $\gamma$-profiles used in
    this paper.  Solid curves show the untruncated profiles
    ($\ghost=0.5, \, M=3.33 \times 10^{11}\,M_{\odot}, \,a=2.33\,
    \mathrm{kpc}$ for the bulge, $\gamma_{\mathrm{DM}}=1, \, M=3.33
    \times 10^{12}\,M_{\odot}, \,a=33.3\, \mathrm{kpc}$ for the halo).
    The dotted lines show truncated profiles for the halo (from left
    to right, $a_{DM}'=66.7,\, 133,\, 333$ kpc; 133 kpc is used for
    the host halos in the simulations), while the dot-dot-dot-dash
    lines show truncated profiles for the bulge (from left to right,
    $a_{\star}'=33.3, \, 83.3,\, 167$ kpc; 167 kpc is used for the
    host bulges in the simulations).  Note that as $a' \rightarrow a$
    (leftmost dotted curve), the inner density amplitude drops but the
    slope remains the same.  \emph{Right:} Stability test of a
    standard satellite for our simulations.  Plotted are the initial
    density profiles (solid curves) for the stellar bulge and dark
    matter halo, as well as the evolved profiles at t = 1 Gyr (dashed
    curves).  The force softening for this run is $\epsilon$ = 33 pc.
    The model is stable on all scales larger than $\sim 2.5\,
    \epsilon$.}
  \label{fig:gammaModels}
\end{figure}

\subsection{Untruncated Models}

\citet[Appendix B]{dehnen1993} gives several useful formulae for
numerically computing quantities such as surface brightness profiles and
projected velocity dispersions for standard $\gamma$ models.  His integral
transforms can be readily extended to both full and truncated two-component
$\gamma$ models, as we show here. 
We first consider the untruncated models composed of three components (dark
matter, stars, and a central black hole), with
$\rho_{\mathrm{DM}}=\rho(\gamma_{\mathrm{DM}}, M_{\mathrm{DM}},
a_{\mathrm{DM}}; r)$ and $\rho_{\star}=\rho(\gamma_{\star}, M_{\star},
a_{\star}; r)$; we set Newton's constant $G$ to unity in the calculations
that follow.

Since the spatial structure of each component is unaffected by the other
component, the formulae in \citet{dehnen1993} for surface density and
cumulative surface density are valid for each component of the untruncated
two-component models.  Computing the projected velocity dispersion is
somewhat more complicated, as the velocity structure of a system is
determined by the total gravitational potential.

The projected stellar velocity dispersion $\sigma_{p,\star}(R)$ as a
function of projected radius $R$ can be computed from
the Jeans equation: 
\begin{eqnarray}
  \sigma_{p,\star}^2(R) & = & \frac{2}{\Sigma_{\star}(R)} \int_R^{\infty}
  \rho_{\star} \sigma_r^2(r) \frac{r}{(r^2-R^2)^{1/2}} \, dr \\
  &=& \frac{2}{\Sigma_{\star}(R)} \int_R^{\infty} \frac{\rho_{\star}
  M_{\mathrm{tot}}(r)} {r^2} (r^2-R^2)^{1/2} \,dr 
\end{eqnarray}
\citep[c.f.][eqns. 27-29]{tremaine1994}, where $\sigma_r^2(r)$ is the
stellar radial velocity dispersion.  The total mass profile can be split
into three parts,
$M_{\mathrm{tot}}(r)=M_{\star}(r)+M_{\mathrm{DM}}(r)+\mbh(r)$, meaning the
integral can be split into three pieces:
$\sigma_{p,\star}^2(r)=s_{p,\star}^2(r)+ s_{p,\mathrm{DM}}^2(r)+
s_{p,BH}^2(r)$.  The first term 
is the projected dispersion for the one-component case (e.g., Dehnen's
eqn.~18), though a singularity in the integrand means it can be much more
easily computed using his eqn.~B2.  The third term can be mapped into an
integral similar to that of the first term
\citep[eqns.~49-51]{tremaine1994} and then computed using Dehnen's
eqn.~B2.  This leaves only one piece to be computed:
\begin{equation}
s_{p,\mathrm{DM}}^2(r)=\frac{2}{\Sigma_{\star}(R)} \int_R^{\infty} \frac{\rho_{\star}
  M_{\mathrm{DM}}(r)} {r^2} (r^2-R^2)^{1/2} \,dr \,.
\label{eqn:sigp_cross}
\end{equation}
This term is difficult to compute numerically because of the limits of
integration.  Following Dehnen, we make the substitution
\begin{equation}
t^2 \equiv \frac{r}{r+a_{\star}}(s+1)-s \,,
\end{equation}
with $s \equiv R/a_{\star}$, and take $a_{\star}=1=M_{\star}$.  
Eqn.~\ref{eqn:sigp_cross} then becomes
\begin{eqnarray}
s_{p,\mathrm{DM}}^2(R) &=& 
\frac{3-\gamma_{\star}}{\pi \Sigma_{\star}(R)}
\frac{M_{\mathrm{DM}} M_{\star}}{a_{\star}^3}
(1+s)^{-(5/2-\gamma_{\star})} \nonumber \\
&\times&  \int_0^1 (1-t^2)^3(t^2+s)^{1-\gamma_{\star}-\gamma_{DM}} \,
[t^2(1-a_{DM})+s+a_{DM}]^{\gamma_{DM}-3}\, t^2\,\sqrt{2s+t^2(1-s)}\, dt \,,
\end{eqnarray}
and numerical integration is straightforward.

\subsection{Truncated Models}
As described in the main text, we truncate the density profiles of our
models by subtracting off a second $\gamma$ profile: assuming
the desired density profile is $\rho_i(\gamma, M, a; r)$, we take the
truncated density profile to be
\begin{equation}
\rho_i(r)=\rho_i(\gamma, M, a; r) - \rho_i(\gamma', M', a'; r) \,.
\end{equation}
In order to preserve the same inner density structure, we require that
$\gamma'=\gamma$, $M' \le M$, and $a' \gg a$.  By varying $M'$
and $a'$, we can control the truncation.
The maximum radius for one component in these models is given by
\begin{equation}
  \label{eqn:rmax}
  r_{\mathrm{max}}=a \frac{\Delta-1} {1-\frac{a}{a'} \Delta}  
\end{equation}
where
\begin{equation}
  \label{eqn:rmax_delta}
  \Delta \equiv \left[\frac{M} {M'} \left(\frac{a'} 
      {a} \right)^{3-\gamma} \right]^{1/(4-\gamma)} \,.
\end{equation}
The left panel of Fig.~\ref{fig:gammaModels} shows the profiles, both
untruncated and truncated, for the dark matter halos and bulges used in
several of the simulations presented in this paper.  The numerical
stability of one of the satellite models is demonstrated in the right panel
of Fig.~\ref{fig:gammaModels}: there is no evolution in the density profile
for scales $\ga 2.5 \, \epsilon$.  The host models are equally stable, both
with and without black holes.

Since the truncated models do not have infinite extent, many of the
integration limits change.  For example, the velocity dispersion has
$\int_r^{\infty} \rightarrow \int_r^{r_{\mathrm{max}}}$, where
$r_{\mathrm{max}}$ is given by eqn.~\ref{eqn:rmax}.  Projected quantities
therefore have an upper limit of $r_{\mathrm{max}}$ rather than $\infty$,
or equivalently, $t_{\mathrm{max}}=t(r_{\mathrm{max}})$ rather than
$t_{\mathrm{max}}=1$.  The calculations are also changed by the addition of
the second profile, which essentially serves as two additional mass
components.  For example, now
$M_{\mathrm{tot}} \rightarrow
M_{\star}-M_{\star '}+M_{\mathrm{DM}}-M_{\mathrm{DM} '}+\mbh$ and 
$\rho_{\star}(r) \rightarrow [\rho_{\star}(r)-\rho_{\star '}(r)]$ (and
similarly for $\rho_{\mathrm{DM}}$).
Velocity dispersions can be computed substituting these relations into the
appropriate equations given above.

The surface brightness for a truncated model is similarly given by
\begin{equation}
\Sigma_{\star}(R) = 2 \int_R^{r_{\mathrm{max}}} 
\frac{\rho(r)}{(r^2-R^2)^{1/2}} \, r\, dr \,
\end{equation}
so introducing a truncation for the bulge in the form of
$\rho_{\star '}(r)=\rho(\gamma_{\star '}, M_{\star '}, a_{\star '}; r)$
is simply a linear operation and 
\begin{equation}
\Sigma_{\star}(R)=\Sigma_{\star}(R)-\Sigma_{\star '}(R) \,.
\end{equation}

\bibliographystyle{apj}
\bibliography{satellites_BH}

\begin{thebibliography}{85}
\expandafter\ifx\csname natexlab\endcsname\relax\def\natexlab#1{#1}\fi

\bibitem[{{Balcells} \& {Quinn}(1990)}]{balcells1990}
{Balcells}, M., \& {Quinn}, P.~J. 1990, \apj, 361, 381

\bibitem[{{Begelman} {et~al.}(1980){Begelman}, {Blandford}, \&
  {Rees}}]{begelman1980}
{Begelman}, M.~C., {Blandford}, R.~D., \& {Rees}, M.~J. 1980, \nat, 287, 307

\bibitem[{{Bender} {et~al.}(1992){Bender}, {Burstein}, \& {Faber}}]{bender1992}
{Bender}, R., {Burstein}, D., \& {Faber}, S.~M. 1992, \apj, 399, 462

\bibitem[{{Benson}(2005)}]{benson2005}
{Benson}, A.~J. 2005, \mnras, 358, 551

\bibitem[{{Binney} \& {Tremaine}(1987)}]{binney1987}
{Binney}, J., \& {Tremaine}, S. 1987, {Galactic Dynamics} (Princeton, NJ,
  Princeton University Press)

\bibitem[{{Bower} {et~al.}(1998){Bower}, {Kodama}, \& {Terlevich}}]{bower1998}
{Bower}, R.~G., {Kodama}, T., \& {Terlevich}, A. 1998, \mnras, 299, 1193

\bibitem[{{Boylan-Kolchin} {et~al.}(2005){Boylan-Kolchin}, {Ma}, \&
  {Quataert}}]{boylan-kolchin2005}
{Boylan-Kolchin}, M., {Ma}, C.-P., \& {Quataert}, E. 2005, \mnras, 362, 184

\bibitem[{{Boylan-Kolchin} {et~al.}(2006){Boylan-Kolchin}, {Ma}, \&
  {Quataert}}]{boylan-kolchin2006}
---. 2006, \mnras, 369, 1081

\bibitem[{{Carlberg}(1984)}]{carlberg1984}
{Carlberg}, R.~G. 1984, \apj, 286, 403

\bibitem[{{Carollo} {et~al.}(1993){Carollo}, {Danziger}, \&
  {Buson}}]{carollo1993}
{Carollo}, C.~M., {Danziger}, I.~J., \& {Buson}, L. 1993, \mnras, 265, 553

\bibitem[{{Chatterjee} {et~al.}(2002){Chatterjee}, {Hernquist}, \&
  {Loeb}}]{chatterjee2002}
{Chatterjee}, P., {Hernquist}, L., \& {Loeb}, A. 2002, \prl, 88, 121103

\bibitem[{{Cox} {et~al.}(2006){Cox}, {Dutta}, {Di Matteo}, {Hernquist},
  {Hopkins}, {Robertson}, \& {Springel}}]{cox2006}
{Cox}, T.~J., {Dutta}, S.~N., {Di Matteo}, T., {Hernquist}, L., {Hopkins},
  P.~F., {Robertson}, B., \& {Springel}, V. 2006, {astro-ph/0607446}

\bibitem[{{Davies} {et~al.}(1983){Davies}, {Efstathiou}, {Fall}, {Illingworth},
  \& {Schechter}}]{davies1983}
{Davies}, R.~L., {Efstathiou}, G., {Fall}, S.~M., {Illingworth}, G., \&
  {Schechter}, P.~L. 1983, \apj, 266, 41

\bibitem[{{de Vaucouleurs}(1948)}]{de-vaucouleurs1948}
{de Vaucouleurs}, G. 1948, Annales d'Astrophysique, 11, 247

\bibitem[{{de Zeeuw} \& {Franx}(1991)}]{de-zeeuw1991}
{de Zeeuw}, T., \& {Franx}, M. 1991, \araa, 29, 239

\bibitem[{{Dehnen}(1993)}]{dehnen1993}
{Dehnen}, W. 1993, \mnras, 265, 250

\bibitem[{{Dekel} {et~al.}(2003){Dekel}, {Devor}, \& {Hetzroni}}]{dekel2003}
{Dekel}, A., {Devor}, J., \& {Hetzroni}, G. 2003, \mnras, 341, 326

\bibitem[{{Djorgovski} \& {Davis}(1987)}]{djorgovski1987}
{Djorgovski}, S., \& {Davis}, M. 1987, \apj, 313, 59

\bibitem[{{Dressler} {et~al.}(1987){Dressler}, {Lynden-Bell}, {Burstein},
  {Davies}, {Faber}, {Terlevich}, \& {Wegner}}]{dressler1987}
{Dressler}, A., {Lynden-Bell}, D., {Burstein}, D., {Davies}, R.~L., {Faber},
  S.~M., {Terlevich}, R., \& {Wegner}, G. 1987, \apj, 313, 42

\bibitem[{{Ebisuzaki} {et~al.}(1991){Ebisuzaki}, {Makino}, \&
  {Okumura}}]{ebisuzaki1991}
{Ebisuzaki}, T., {Makino}, J., \& {Okumura}, S.~K. 1991, \nat, 354, 212

\bibitem[{{Faber} \& {Jackson}(1976)}]{faber1976}
{Faber}, S.~M., \& {Jackson}, R.~E. 1976, \apj, 204, 668

\bibitem[{{Faber} {et~al.}(1997){Faber}, {Tremaine}, {Ajhar}, {Byun},
  {Dressler}, {Gebhardt}, {Grillmair}, {Kormendy}, {Lauer}, \&
  {Richstone}}]{faber1997}
{Faber}, S.~M. {et~al.} 1997, \aj, 114, 1771

\bibitem[{{Ferrarese} {et~al.}(2006){Ferrarese}, {C{\^o}t{\'e}}, {Jord{\'a}n},
  {Peng}, {Blakeslee}, {Piatek}, {Mei}, {Merritt}, {Milosavljevi{\'c}},
  {Tonry}, \& {West}}]{ferrarese2006}
{Ferrarese}, L. {et~al.} 2006, \apjs, 164, 334

\bibitem[{{Ferrarese} \& {Merritt}(2000)}]{ferrarese2000}
{Ferrarese}, L., \& {Merritt}, D. 2000, \apjl, 539, L9

\bibitem[{{Gao} {et~al.}(2004){Gao}, {De Lucia}, {White}, \&
  {Jenkins}}]{gao2004a}
{Gao}, L., {De Lucia}, G., {White}, S.~D.~M., \& {Jenkins}, A. 2004, \mnras,
  352, L1

\bibitem[{{Gebhardt} {et~al.}(2000){Gebhardt}, {Bender}, {Bower}, {Dressler},
  {Faber}, {Filippenko}, {Green}, {Grillmair}, {Ho}, {Kormendy}, {Lauer},
  {Magorrian}, {Pinkney}, {Richstone}, \& {Tremaine}}]{gebhardt2000}
{Gebhardt}, K. {et~al.} 2000, \apjl, 539, L13

\bibitem[{{Gebhardt} {et~al.}(2003){Gebhardt}, {Richstone}, {Tremaine},
  {Lauer}, {Bender}, {Bower}, {Dressler}, {Faber}, {Filippenko}, {Green},
  {Grillmair}, {Ho}, {Kormendy}, {Magorrian}, \& {Pinkney}}]{gebhardt2003}
---. 2003, \apj, 583, 92

\bibitem[{{Genzel} {et~al.}(2001){Genzel}, {Tacconi}, {Rigopoulou}, {Lutz}, \&
  {Tecza}}]{genzel2001}
{Genzel}, R., {Tacconi}, L.~J., {Rigopoulou}, D., {Lutz}, D., \& {Tecza}, M.
  2001, \apj, 563, 527

\bibitem[{{Gnedin}(2003)}]{gnedin2003}
{Gnedin}, O.~Y. 2003, \apj, 589, 752

\bibitem[{{Gnedin} {et~al.}(1999){Gnedin}, {Hernquist}, \&
  {Ostriker}}]{gnedin1999}
{Gnedin}, O.~Y., {Hernquist}, L., \& {Ostriker}, J.~P. 1999, \apj, 514, 109

\bibitem[{{Gnedin} \& {Ostriker}(1999)}]{gnedin1999a}
{Gnedin}, O.~Y., \& {Ostriker}, J.~P. 1999, \apj, 513, 626

\bibitem[{{Gregg} \& {West}(1998)}]{gregg1998}
{Gregg}, M.~D., \& {West}, M.~J. 1998, \nat, 396, 549

\bibitem[{{H{\"a}ring} \& {Rix}(2004)}]{haring2004}
{H{\"a}ring}, N., \& {Rix}, H. 2004, \apjl, 604, L89

\bibitem[{{Hausman} \& {Ostriker}(1978)}]{hausman1978}
{Hausman}, M.~A., \& {Ostriker}, J.~P. 1978, \apj, 224, 320

\bibitem[{{Hayashi} {et~al.}(2003){Hayashi}, {Navarro}, {Taylor}, {Stadel}, \&
  {Quinn}}]{hayashi2003}
{Hayashi}, E., {Navarro}, J.~F., {Taylor}, J.~E., {Stadel}, J., \& {Quinn}, T.
  2003, \apj, 584, 541

\bibitem[{{Hernquist}(1990)}]{hernquist1990}
{Hernquist}, L. 1990, \apj, 356, 359

\bibitem[{{Hernquist} \& {Barnes}(1991)}]{hernquist1991}
{Hernquist}, L., \& {Barnes}, J.~E. 1991, \nat, 354, 210

\bibitem[{{Hogg} {et~al.}(2004){Hogg}, {Blanton}, {Brinchmann}, {Eisenstein},
  {Schlegel}, {Gunn}, {McKay}, {Rix}, {Bahcall}, {Brinkmann}, \&
  {Meiksin}}]{hogg2004}
{Hogg}, D.~W. {et~al.} 2004, \apjl, 601, L29

\bibitem[{{Holley-Bockelmann} \& {Richstone}(2000)}]{holley-bockelmann2000}
{Holley-Bockelmann}, K., \& {Richstone}, D.~O. 2000, \apj, 531, 232

\bibitem[{{Jesseit} {et~al.}(2006){Jesseit}, {Naab}, {Peletier}, \&
  {Burkert}}]{jesseit2006}
{Jesseit}, R., {Naab}, T., {Peletier}, R., \& {Burkert}, A. 2006,
  {astro-ph/0606144}

\bibitem[{{Kauffmann} {et~al.}(2003){Kauffmann}, {Heckman}, {White}, {Charlot},
  {Tremonti}, {Brinchmann}, {Bruzual}, {Peng}, {Seibert}, {Bernardi},
  {Blanton}, {Brinkmann}, {Castander}, {Cs{\'a}bai}, {Fukugita}, {Ivezic},
  {Munn}, {Nichol}, {Padmanabhan}, {Thakar}, {Weinberg}, \&
  {York}}]{kauffmann2003}
{Kauffmann}, G. {et~al.} 2003, \mnras, 341, 33

\bibitem[{{Kazantzidis} {et~al.}(2004){Kazantzidis}, {Mayer}, {Mastropietro},
  {Diemand}, {Stadel}, \& {Moore}}]{kazantzidis2004b}
{Kazantzidis}, S., {Mayer}, L., {Mastropietro}, C., {Diemand}, J., {Stadel},
  J., \& {Moore}, B. 2004, \apj, 608, 663

\bibitem[{{Khochfar} \& {Burkert}(2006)}]{khochfar2006}
{Khochfar}, S., \& {Burkert}, A. 2006, \aap, 445, 403

\bibitem[{{Kormendy}(1984)}]{kormendy1984}
{Kormendy}, J. 1984, \apj, 287, 577

\bibitem[{{Kormendy} \& {Bender}(1996)}]{kormendy1996}
{Kormendy}, J., \& {Bender}, R. 1996, \apjl, 464, L119

\bibitem[{{Kravtsov} {et~al.}(2004){Kravtsov}, {Gnedin}, \&
  {Klypin}}]{kravtsov2004}
{Kravtsov}, A.~V., {Gnedin}, O.~Y., \& {Klypin}, A.~A. 2004, \apj, 609, 482

\bibitem[{{Larson}(1974)}]{larson1974}
{Larson}, R.~B. 1974, \mnras, 166, 585

\bibitem[{{Lauer}(1988)}]{lauer1988}
{Lauer}, T.~R. 1988, \apj, 325, 49

\bibitem[{{Lauer} {et~al.}(1995){Lauer}, {Ajhar}, {Byun}, {Dressler}, {Faber},
  {Grillmair}, {Kormendy}, {Richstone}, \& {Tremaine}}]{lauer1995}
{Lauer}, T.~R. {et~al.} 1995, \aj, 110, 2622

\bibitem[{{Lauer} {et~al.}(2005){Lauer}, {Faber}, {Gebhardt}, {Richstone},
  {Tremaine}, {Ajhar}, {Aller}, {Bender}, {Dressler}, {Filippenko}, {Green},
  {Grillmair}, {Ho}, {Kormendy}, {Magorrian}, {Pinkney}, \&
  {Siopis}}]{lauer2005}
---. 2005, \aj, 129, 2138

\bibitem[{{Lauer} {et~al.}(2006){Lauer}, {Faber}, {Richstone}, {Gebhardt},
  {Tremaine}, {Postman}, {Dressler}, {Aller}, {Filippenko}, {Green}, {Ho},
  {Kormendy}, {Magorrian}, \& {Pinkney}}]{lauer2006}
---. 2006, {astro-ph/0606739}

\bibitem[{{Lauer} {et~al.}(2002){Lauer}, {Gebhardt}, {Richstone}, {Tremaine},
  {Bender}, {Bower}, {Dressler}, {Faber}, {Filippenko}, {Green}, {Grillmair},
  {Ho}, {Kormendy}, {Magorrian}, {Pinkney}, {Laine}, {Postman}, \& {van der
  Marel}}]{lauer2002}
---. 2002, \aj, 124, 1975

\bibitem[{{Ma} \& {Boylan-Kolchin}(2004)}]{ma2004a}
{Ma}, C., \& {Boylan-Kolchin}, M. 2004, \prl, 93, 021301

\bibitem[{{Magorrian} {et~al.}(1998){Magorrian}, {Tremaine}, {Richstone},
  {Bender}, {Bower}, {Dressler}, {Faber}, {Gebhardt}, {Green}, {Grillmair},
  {Kormendy}, \& {Lauer}}]{magorrian1998}
{Magorrian}, J. {et~al.} 1998, \aj, 115, 2285

\bibitem[{{McDermid} {et~al.}(2006){McDermid}, {Bacon}, {Kuntschner},
  {Emsellem}, {Shapiro}, {Bureau}, {Cappellari}, {Davies},
  {Falc{\'o}n-Barroso}, {Krajnovi{\'c}}, {Peletier}, {Sarzi}, \& {de
  Zeeuw}}]{mcdermid2006}
{McDermid}, R.~M. {et~al.} 2006, New Astronomy Review, 49, 521

\bibitem[{{Merritt}(1984)}]{merritt1984}
{Merritt}, D. 1984, \apj, 276, 26

\bibitem[{{Merritt}(1985)}]{merritt1985}
---. 1985, \apj, 289, 18

\bibitem[{{Merritt}(2006)}]{merritt2006a}
---. 2006, {astro-ph/0603439}

\bibitem[{{Merritt} \& {Cruz}(2001)}]{merritt2001}
{Merritt}, D., \& {Cruz}, F. 2001, \apjl, 551, L41

\bibitem[{{Mihos} {et~al.}(2005){Mihos}, {Harding}, {Feldmeier}, \&
  {Morrison}}]{mihos2005}
{Mihos}, J.~C., {Harding}, P., {Feldmeier}, J., \& {Morrison}, H. 2005, \apjl,
  631, L41

\bibitem[{{Milosavljevi{\'c}} \& {Merritt}(2001)}]{milosavljevic2001}
{Milosavljevi{\'c}}, M., \& {Merritt}, D. 2001, \apj, 563, 34

\bibitem[{{Moore} {et~al.}(1996){Moore}, {Katz}, {Lake}, {Dressler}, \&
  {Oemler}}]{moore1996}
{Moore}, B., {Katz}, N., {Lake}, G., {Dressler}, A., \& {Oemler}, Jr., A. 1996,
  \nat, 379, 613

\bibitem[{{Navarro} {et~al.}(1997){Navarro}, {Frenk}, \& {White}}]{navarro1997}
{Navarro}, J.~F., {Frenk}, C.~S., \& {White}, S.~D.~M. 1997, \apj, 490, 493

\bibitem[{{Ostriker} {et~al.}(1972){Ostriker}, {Spitzer}, \&
  {Chevalier}}]{ostriker1972}
{Ostriker}, J.~P., {Spitzer}, L.~J., \& {Chevalier}, R.~A. 1972, \apjl, 176,
  L51

\bibitem[{{Ostriker} \& {Tremaine}(1975)}]{ostriker1975}
{Ostriker}, J.~P., \& {Tremaine}, S.~D. 1975, \apjl, 202, L113

\bibitem[{{Padmanabhan} {et~al.}(2004){Padmanabhan}, {Seljak}, {Strauss},
  {Blanton}, {Kauffmann}, {Schlegel}, {Tremonti}, {Bahcall}, {Bernardi},
  {Brinkmann}, {Fukugita}, \& {Ivezi{\'c}}}]{Padmanabhan2004}
{Padmanabhan}, N. {et~al.} 2004, New Astronomy, 9, 329

\bibitem[{{Quinlan} \& {Hernquist}(1997)}]{quinlan1997}
{Quinlan}, G.~D., \& {Hernquist}, L. 1997, New Astronomy, 2, 533

\bibitem[{{Ravindranath} {et~al.}(2001){Ravindranath}, {Ho}, {Peng},
  {Filippenko}, \& {Sargent}}]{ravindranath2001}
{Ravindranath}, S., {Ho}, L.~C., {Peng}, C.~Y., {Filippenko}, A.~V., \&
  {Sargent}, W.~L.~W. 2001, \aj, 122, 653

\bibitem[{{Rudick} {et~al.}(2006){Rudick}, {Mihos}, \& {McBride}}]{rudick2006}
{Rudick}, C.~S., {Mihos}, J.~C., \& {McBride}, C. 2006, {astro-ph/0605603}

\bibitem[{{Shen} {et~al.}(2003){Shen}, {Mo}, {White}, {Blanton}, {Kauffmann},
  {Voges}, {Brinkmann}, \& {Csabai}}]{shen2003}
{Shen}, S., {Mo}, H.~J., {White}, S.~D.~M., {Blanton}, M.~R., {Kauffmann}, G.,
  {Voges}, W., {Brinkmann}, J., \& {Csabai}, I. 2003, \mnras, 343, 978

\bibitem[{{Sommer-Larsen} {et~al.}(2005){Sommer-Larsen}, {Romeo}, \&
  {Portinari}}]{sommer-larsen2005}
{Sommer-Larsen}, J., {Romeo}, A.~D., \& {Portinari}, L. 2005, \mnras, 357, 478

\bibitem[{{Springel}(2005)}]{springel2005}
{Springel}, V. 2005, \mnras, 364, 1105

\bibitem[{{Springel} {et~al.}(2005){Springel}, {Di Matteo}, \&
  {Hernquist}}]{springel2005a}
{Springel}, V., {Di Matteo}, T., \& {Hernquist}, L. 2005, \mnras, 361, 776

\bibitem[{{Taffoni} {et~al.}(2003){Taffoni}, {Mayer}, {Colpi}, \&
  {Governato}}]{taffoni2003}
{Taffoni}, G., {Mayer}, L., {Colpi}, M., \& {Governato}, F. 2003, \mnras, 341,
  434

\bibitem[{{Taylor} \& {Babul}(2001)}]{taylor2001}
{Taylor}, J.~E., \& {Babul}, A. 2001, \apj, 559, 716

\bibitem[{{Tremaine} {et~al.}(1994){Tremaine}, {Richstone}, {Byun}, {Dressler},
  {Faber}, {Grillmair}, {Kormendy}, \& {Lauer}}]{tremaine1994}
{Tremaine}, S., {Richstone}, D.~O., {Byun}, Y., {Dressler}, A., {Faber}, S.~M.,
  {Grillmair}, C., {Kormendy}, J., \& {Lauer}, T.~R. 1994, \aj, 107, 634

\bibitem[{{Velazquez} \& {White}(1999)}]{velazquez1999}
{Velazquez}, H., \& {White}, S.~D.~M. 1999, \mnras, 304, 254

\bibitem[{{Wang} {et~al.}(2006){Wang}, {Li}, {Kauffmann}, \& {De
  Lucia}}]{wang2006}
{Wang}, L., {Li}, C., {Kauffmann}, G., \& {De Lucia}, G. 2006,
  {astro-ph/0603546}, 785

\bibitem[{{Weinberg}(1994)}]{weinberg1994}
{Weinberg}, M.~D. 1994, \aj, 108, 1403

\bibitem[{{Weinberg}(1997)}]{weinberg1997}
---. 1997, \apj, 478, 435

\bibitem[{{White}(1980)}]{white1980}
{White}, S.~D.~M. 1980, \mnras, 191, 1

\bibitem[{{White}(1983)}]{white1983}
---. 1983, \apj, 274, 53

\bibitem[{{Willman} {et~al.}(2004){Willman}, {Governato}, {Wadsley}, \&
  {Quinn}}]{willman2004}
{Willman}, B., {Governato}, F., {Wadsley}, J., \& {Quinn}, T. 2004, \mnras,
  355, 159

\bibitem[{{York} {et~al.}(2000){York}, {Adelman}, {Anderson}, {Anderson},
  {Annis}, {Bahcall}, {Bakken}, {Barkhouser}, {Bastian}, {Berman}, {Boroski},
  {Bracker}, {Briegel}, {Briggs}, {Brinkmann}, {Brunner}, {Burles}, {Carey},
  {Carr}, {Castander}, {Chen}, {Colestock}, {Connolly}, {Crocker}, {Csabai},
  {Czarapata}, {Davis}, {Doi}, {Dombeck}, {Eisenstein}, {Ellman}, {Elms},
  {Evans}, {Fan}, {Federwitz}, {Fiscelli}, {Friedman}, {Frieman}, {Fukugita},
  {Gillespie}, {Gunn}, {Gurbani}, {de Haas}, {Haldeman}, {Harris}, {Hayes},
  {Heckman}, {Hennessy}, {Hindsley}, {Holm}, {Holmgren}, {Huang}, {Hull},
  {Husby}, {Ichikawa}, {Ichikawa}, {Ivezi{\'c}}, {Kent}, {Kim}, {Kinney},
  {Klaene}, {Kleinman}, {Kleinman}, {Knapp}, {Korienek}, {Kron}, {Kunszt},
  {Lamb}, {Lee}, {Leger}, {Limmongkol}, {Lindenmeyer}, {Long}, {Loomis},
  {Loveday}, {Lucinio}, {Lupton}, {MacKinnon}, {Mannery}, {Mantsch}, {Margon},
  {McGehee}, {McKay}, {Meiksin}, {Merelli}, {Monet}, {Munn}, {Narayanan},
  {Nash}, {Neilsen}, {Neswold}, {Newberg}, {Nichol}, {Nicinski}, {Nonino},
  {Okada}, {Okamura}, {Ostriker}, {Owen}, {Pauls}, {Peoples}, {Peterson},
  {Petravick}, {Pier}, {Pope}, {Pordes}, {Prosapio}, {Rechenmacher}, {Quinn},
  {Richards}, {Richmond}, {Rivetta}, {Rockosi}, {Ruthmansdorfer}, {Sandford},
  {Schlegel}, {Schneider}, {Sekiguchi}, {Sergey}, {Shimasaku}, {Siegmund},
  {Smee}, {Smith}, {Snedden}, {Stone}, {Stoughton}, {Strauss}, {Stubbs},
  {SubbaRao}, {Szalay}, {Szapudi}, {Szokoly}, {Thakar}, {Tremonti}, {Tucker},
  {Uomoto}, {Vanden Berk}, {Vogeley}, {Waddell}, {Wang}, {Watanabe},
  {Weinberg}, {Yanny}, \& {Yasuda}}]{york2000}
{York}, D.~G. {et~al.} 2000, \aj, 120, 1579

\bibitem[{{Zibetti} {et~al.}(2005){Zibetti}, {White}, {Schneider}, \&
  {Brinkmann}}]{zibetti2005}
{Zibetti}, S., {White}, S.~D.~M., {Schneider}, D.~P., \& {Brinkmann}, J. 2005,
  \mnras, 358, 949

\end{thebibliography}

\end{document}